\documentclass[%
 reprint,
superscriptaddress,
nofootinbib,
 showpacs,amsmath,amssymb,
 aps,
floatfix
]{revtex4-2}

\usepackage{graphicx}% Include figure files
\usepackage{dcolumn}% Align table columns on decimal point
\usepackage{bm}% bold math
\usepackage[colorlinks = true,
            linkcolor = blue,
            urlcolor  = blue,
            citecolor = blue,
            anchorcolor = blue]{hyperref}% add hypertext capabilities
\usepackage{lipsum}
\usepackage{comment}
\usepackage{siunitx}
\usepackage[symbol]{footmisc} % use symbols for footnotes rather than numbers
\usepackage{threeparttable}

\usepackage{lineno}
\DeclareSIUnit{\litre}{\ell}
\DeclareSIUnit{\gauss}{Gs}

\usepackage{xcolor}
\newcommand{\td}[1]{\textcolor{magenta}{(TD: #1)}}
\newcommand{\red}[1]{\textcolor{red}{#1}}
\DeclareUrlCommand\pkg{\urlstyle{tt}}

\begin{document}

\preprint{APS/123-QED}
\draft
\title{Design of a Resonator Array Axion Haloscope \\ and its Projected Sensitivity from 2 to 4 GHz}

\author{T. A. Dyson}\email[Correspondence to: tdyson@stanford.edu]{} %x
\affiliation{Stanford University, Stanford, CA 94305, USA}
%\affiliation{Kavli Institute for Particle Astrophysics and Cosmology, Menlo Park, CA 94025, USA}

\author{M. H. Awida}
\affiliation{Fermi National Accelerator Laboratory, Batavia, Illinois 60510, USA}

\author{C. Bartram} %x
\affiliation{SLAC National Accelerator Laboratory, Menlo Park, CA 94025, USA}
%\affiliation{Kavli Institute for Particle Astrophysics and Cosmology, Menlo Park, CA 94025, USA}

\author{C. Boutan} %x
\affiliation{Pacific Northwest National Laboratory, Richland, Washington 99354, USA}

\author{T. Braine} %x
\affiliation{Pacific Northwest National Laboratory, Richland, Washington 99354, USA}

\author{J. H. Buckley} %x
\affiliation{Washington University, St. Louis, Missouri 63130, USA}

\author{T. J. Caligiure}
\affiliation{University of Florida, Gainesville, Florida 32611, USA}

\author{M. Carhart} %x
\affiliation{University of Washington, Seattle, Washington 98195, USA}

\author{G. Carosi} % x
\affiliation{Lawrence Livermore National Laboratory, Livermore, California 94550, USA}

\author{A. S. Chou} % x
\affiliation{Fermi National Accelerator Laboratory, Batavia, Illinois 60510, USA}

\author{N. Christman} % x Nicholas Christman
\affiliation{Pacific Northwest National Laboratory, Richland, Washington 99354, USA}

\author{John Clarke}
\affiliation{University of California, Berkeley, California 94720, USA}

\author{E. J. Daw} %x
\affiliation{University of Sheffield, Sheffield S10 2TN, UK}

\author{N. Du} % x
\affiliation{Lawrence Livermore National Laboratory, Livermore, California 94550, USA}

\author{L. D. Duffy} %x
\affiliation{Los Alamos National Laboratory, Los Alamos, New Mexico 87545, USA}

\author{S. Durham}
\affiliation{Lawrence Livermore National Laboratory, Livermore, California 94550, USA}

\author{K. Enzian} %x
\affiliation{University of Washington, Seattle, Washington 98195, USA}

\author{G. Flower} %x
\affiliation{University of Western Australia, Perth, Western Australia 6009, Australia}

\author{C. Gaikwad}
\affiliation{Washington University, St. Louis, Missouri 63130, USA}

\author{J. R. Gleason}
\affiliation{University of Florida, Gainesville, Florida 32611, USA}

\author{C. Goodman} %x
\affiliation{University of Washington, Seattle, Washington 98195, USA}

\author{M. Goryachev} %x
\affiliation{University of Western Australia, Perth, Western Australia 6009, Australia}

\author{C. Hanretty} %x
\affiliation{University of Washington, Seattle, Washington 98195, USA}

\author{M. T. Hatzon} %x
\affiliation{University of Western Australia, Perth, Western Australia 6009, Australia}

\author{A. T. Hipp} %x
\affiliation{University of Florida, Gainesville, Florida 32611, USA}

\author{J. Hoffman} %x
\affiliation{Washington University, St. Louis, Missouri 63130, USA}

\author{M. Hollister}
\affiliation{Fermi National Accelerator Laboratory, Batavia, Illinois 60510, USA}

\author{M. Jones} %x
\affiliation{Pacific Northwest National Laboratory, Richland, Washington 99354, USA}

\author{R. Khatiwada} % x
\affiliation{Illinois Institute of Technology, Chicago, Illinois 60616, USA}
\affiliation{Fermi National Accelerator Laboratory, Batavia, Illinois 60510, USA}

\author{O. Kim} %x
\affiliation{University of Washington, Seattle, Washington 98195, USA}

\author{S. Knirck} %x
\affiliation{Harvard University, Cambridge, Massachusetts 02138, USA}

\author{D. Kowsari} %x
\affiliation{Washington University, St. Louis, Missouri 63130, USA}

\author{C. Kuo} %x
\affiliation{Stanford University, Stanford, CA 94305, USA}
\affiliation{SLAC National Accelerator Laboratory, Menlo Park, CA 94025, USA}
%\affiliation{Kavli Institute for Particle Astrophysics and Cosmology, Menlo Park, CA 94025, USA}

\author{E. Lentz} %x
\affiliation{Pacific Northwest National Laboratory, Richland, Washington 99354, USA}

\author{M. J. Lynn} %x
\affiliation{Harvard University, Cambridge, Massachusetts 02138, USA}

\author{B. T. McAllister} %x
\affiliation{Swinburne University of Technology, Melbourne, Victoria 3122, Australia}

\author{D. V. Mitchell}
\affiliation{Fermi National Accelerator Laboratory, Batavia, Illinois 60510, USA}

\author{C. Mostyn} %x
\affiliation{University of Sheffield, Sheffield S10 2TN, UK}

\author{K. W. Murch} %x
\affiliation{Washington University, St. Louis, Missouri 63130, USA}

\author{T. N. L. Nguyen}
\affiliation{University of Florida, Gainesville, Florida 32611, USA}

\author{N. S.~Oblath} %x
\affiliation{Pacific Northwest National Laboratory, Richland, Washington 99354, USA}

\author{E. C. I. Paterson} % Emma Paterson
\affiliation{University of Western Australia, Perth, Western Australia 6009, Australia}

\author{M. G. Perry} %x
\affiliation{University of Sheffield, Sheffield S10 2TN, UK}

\author{A. Quiskamp} %x
\affiliation{University of Western Australia, Perth, Western Australia 6009, Australia}

\author{S. Ruppert} %x
\affiliation{Stanford University, Stanford, CA 94305, USA}

\author{B. J. Russell} %x
\affiliation{Washington University, St. Louis, Missouri 63130, USA}

\author{G. Rybka} %x
\affiliation{University of Washington, Seattle, Washington 98195, USA}

\author{P. Sikivie} % x
\affiliation{University of Florida, Gainesville, Florida 32611, USA}

\author{R. A. Singh} % x Raj Aryan Singh
\affiliation{Swinburne University of Technology, Melbourne, Victoria 3122, Australia}

\author{J. Sinnis} %x
\affiliation{University of Washington, Seattle, Washington 98195, USA}

\author{M. E. Solano}
\affiliation{University of Florida, Gainesville, Florida 32611, USA}

\author{A. Sonnenschein} %x
\affiliation{Fermi National Accelerator Laboratory, Batavia, Illinois 60510, USA}

\author{N. S. Sullivan} % x
\affiliation{University of Florida, Gainesville, Florida 32611, USA}

\author{D. B. Tanner} % x
\affiliation{University of Florida, Gainesville, Florida 32611, USA}

\author{M.~S. Taubman} %x
\affiliation{Pacific Northwest National Laboratory, Richland, Washington 99354, USA}

\author{J. Tedeschi} % x Jonathan Tedeschi
\affiliation{Pacific Northwest National Laboratory, Richland, Washington 99354, USA}

\author{R. J. Tesarek} %x
\affiliation{Fermi National Accelerator Laboratory, Batavia, Illinois 60510, USA}

\author{M. E. Tobar} %x
\affiliation{University of Western Australia, Perth, Western Australia 6009, Australia}

\author{M. O. Withers} %x
\affiliation{Stanford University, Stanford, CA 94305, USA}
%\affiliation{Kavli Institute for Particle Astrophysics and Cosmology, Menlo Park, CA 94025, USA}

\author{A. Yi} %x
\affiliation{SLAC National Accelerator Laboratory, Menlo Park, CA 94025, USA}
%\affiliation{Kavli Institute for Particle Astrophysics and Cosmology, Menlo Park, CA 94025, USA}

\author{B. F. Zemenu} %x
\affiliation{Stanford University, Stanford, CA 94305, USA}
%\affiliation{Kavli Institute for Particle Astrophysics and Cosmology, Menlo Park, CA 94025, USA}

\author{D. Zhang} %x
\affiliation{University of Washington, Seattle, Washington 98195, USA}

\author{Z. C. Zhao} %x
\affiliation{University of Western Australia, Perth, Western Australia 6009, Australia}

\author{K. Zheng} %x
\affiliation{Washington University, St. Louis, Missouri 63130, USA}

\collaboration{ADMX Collaboration}\noaffiliation

\date{\today}% It is always \today, today,
             %  but any date may be explicitly specified

\begin{abstract}
\noindent
The QCD axion is a dark matter (DM) candidate particle which can also solve the long-standing strong charge-parity problem. The Axion Dark Matter eXperiment (ADMX) searches for resonantly-enhanced conversion of axion DM into photons in a strong magnetic field. Previous versions of the experiment probed most QCD axion DM from \SI{2.7}{\micro\electronvolt} to \SI{4.2}{\micro\electronvolt}, with ongoing and planned runs poised to cover up to \SI{8}{\micro\electronvolt} (\SI{2}{\giga\hertz}). We present the design of a haloscope with 18 cavities \SI{1}{\metre} in length read out using individual low-noise amplifier chains and digital coherent signal combination called ADMX extended frequency range (EFR). We discuss details on magnet and cryogenic infrastructure, cavity design, cold electronics and efficiency of coherent signal combination. We use detailed cavity and electronics simulations, as well as diagnostic data from previous ADMX runs, to estimate frequency-dependent instantaneous scan rate. When paired with a comprehensive study of experiment downtime, a projection of final axion coupling sensitivity is produced. Under the stated assumptions, the projected sensitivity is projected to reach below the DFSZ axion benchmark model from \SI{8}{\micro\electronvolt} to \SI{16}{\micro\electronvolt} (\SI{2}{\giga\hertz} to \SI{4}{\giga\hertz}) in 36 months.
\end{abstract}
% Before reframing:
%The QCD axion is a dark matter (DM) candidate particle which can also solve the long-standing strong charge-parity problem. The Axion Dark Matter eXperiment (ADMX) searches for resonantly-enhanced conversion of axion DM into photons in a strong magnetic field. Previous versions of the experiment probed most QCD axion DM from \SI{2.7}{\micro\electronvolt} to \SI{4.2}{\micro\electronvolt}, with ongoing and planned runs poised to cover up to \SI{8}{\micro\electronvolt} (\SI{2}{\giga\hertz}). We present ADMX extended frequency range (EFR) and evaluate it in terms of its sensitivity to QCD axions in a mass range \SI{8}{\micro\electronvolt} to \SI{16}{\micro\electronvolt} (\SI{2}{\giga\hertz} to \SI{4}{\giga\hertz}). ADMX-EFR consists of an array of 18 cavities \SI{1}{\metre} in length read out using individual low-noise amplifier chains. We discuss details on magnet and cryogenic infrastructure, cavity design, cold electronics and efficiency of coherent signal combination. We use detailed cavity and electronics simulations, as well as data from previous ADMX runs, to estimate frequency-dependent instantaneous scan rate. When paired with a comprehensive study of experiment downtime, a robust projection of final axion coupling sensitivity is produced. Sensitivity to the DFSZ axion benchmark model over an octave in mass range is achieved in an estimated experimental duration of 36 months.

%\keywords{Suggested keywords}%Use showkeys class option if keyword
                              %display desired
\maketitle

%\tableofcontents

\section{\label{sec:introduction}Introduction}

\noindent
The axion is a hypothetical particle introduced to solve the strong CP-problem~\cite{Peccei:1977hh,Wilczek:1977pj,Weinberg:1977ma} --- that the strong interaction preserves charge-parity symmetry, resulting in a vanishingly small electric dipole moment of the neutron~\cite{Harris:1999jx,Baker:2006ts,Pendlebury:2015lrz,Abel:2020gbr}. Axions are prime dark matter (DM) candidates because they have early-universe production mechanisms that allow them to account for all DM~\cite{Preskill:1982cy,Abbott:1982af,Dine:1982ah,PhysRevD.85.105020}.
The axion has a coupling to two photons, $g_{a\gamma\gamma}={\alpha}g_{\gamma}/{{\pi}f_a}$, where $\alpha$ is the fine structure constant, and $f_a$ is the PQ symmetry breaking scale. The dimensionless parameter $g_{\gamma}$ is predicted by two benchmark models to be $g_{\gamma} = -0.97$ (KSVZ model)~\cite{Kim:1979if,Shifman:1979if} and $g_{\gamma} = 0.36$ (DFSZ model)~\cite{Dine:1981rt,Zhitnitsky:1980tq}.

A growing number of experiments aim to detect axion DM directly~\cite{Irastorza:2018dyq,RevModPhys.93.015004,Adams:2022pbo}.
Many experiments use a strong magnetic field to convert axions to photons via the inverse Primakoff effect and an electromagnetic resonator with high quality factor $Q$ to enhance the conversion, as first proposed by Sikivie~\cite{PhysRevLett.51.1415}.
Most notably, DFSZ axions as dark matter have been excluded using resonant microwave cavities for most axion masses $m_a$ of \SI{2.7}{\micro\electronvolt} to \SI{4.2}{\micro\electronvolt}, under an assumption of local $\rho_\mathrm{DM}=\SI{0.45}{\giga\electronvolt\per\centi\meter\cubed}$ by the Axion Dark Matter eXperiment (ADMX)~\cite{ADMX:2018gho,ADMX:2019uok,ADMX:2021nhd}, and from \SI{4.5}{\micro\electronvolt} to \SI{4.6}{\micro\electronvolt} by the Institute for Basic Science's Center for Axion and Precision Physics Research (IBS-CAPP)~\cite{Yi:2022fmn}.
%In this paper we present the conceptual design of ADMX extended frequency range (EFR), targeting $m_a$ from \SI{8}{\micro\electronvolt} to \SI{16}{\micro\electronvolt}. This study is intended to serve as a guide for what is possible to accomplish with a resonator array haloscope at these frequencies, and can serve as a baseline for comparison with future haloscope designs.
In this paper, we present a conceptual design for an 18 cavity array targeting $m_a$ from \SI{8}{\micro\electronvolt} to \SI{16}{\micro\electronvolt} (\SI{2}{\giga\hertz} to \SI{4}{\giga\hertz}). We refer to the design as ADMX extended frequency range (EFR) for convenience throughout the text.

Because DM is known to move with non-relativistic speeds $v_a \lesssim 10^{-3}c$, the energy of incoming axions is dominated by $m_a$. The outgoing photon from Primakoff conversion thus has a frequency $\nu_a = m_a c^2 / h$ and a line width $\Delta\nu_a=\nu_a/Q_a\approx \nu_a/10^{6}$ set by the axions' velocity dispersion~\cite{Irastorza:2018dyq}. Thus, searching for an axion of higher mass requires a cavity with an axion-sensitive mode at higher frequency. The mode with the greatest axion sensitivity in a cylinder is the TM$_{010}$ mode (hereafter referred to as ``the mode''). For the mode to have a higher resonant frequency, the cylinder must have a smaller diameter, drastically reducing detector volume and thus its ultimate sensitivity. ADMX-EFR proposes to surmount this challenge by filling out the available conversion volume (constrained by the dimensions of the solenoid magnet bore) with an array of 18 cylindrical cavities, reading each out individually and combining the signals digitally. A set of synchronously actuated rods keeps the  frequencies of the modes in each cavity equal, and tunes the cavities across their range with instantaneous scan rate $d\nu / dt$ for a given set of experimental parameters and desired sensitivity to DFSZ axion DM, given by
\begin{widetext}
\begin{equation}
\label{eq:instscanspeed}
\begin{aligned}
    \frac{d\nu}{dt}~{\approx}~1.75~\frac{\mathrm{GHz}}{\mathrm{yr}} &\times 
    \left(\frac{B}{\SI{9.1}{\tesla}}\right)^4 & \text{(\ref{sec:magnet}. MAGNET)} \\
    %experimental
    &\times 
    \left(\frac{C}{0.4}\right)^2
    \left(\frac{NV}{18 \times \SI{10.1}{\litre}}\right)^2
    \left(\frac{Q_{\text{L}}}{50{,}000}\right) & \text{(\ref{sec:cavity}.~RESONATOR)} \\
    & \times
    \left(\frac{\beta}{1 + \beta }\right)^2 
    \left(\frac{\SI{800}{\milli\kelvin}}{T_{\text{sys}}}\right)^2 & \text{(\ref{sec:electronics}.~ELECTRONICS)} \\
    & \times \left(\frac{\eta_c}{1}\right)^2  & \text{(\ref{sec:powercomb}.~SIGNAL COMBINATION)}  \\
    % statistics
    &\times \left(\frac{3.5}{\text{SNR}}\right)^2
    % dark matter
    \bigg(\frac{g_{\gamma}}{0.36}\bigg)^4
    \left(\frac{\nu}{\SI{4}{\giga\hertz}}\right)^2
    \left(\frac{\rho_\mathrm{DM}}{0.45\mathrm{~\mathrm{GeV/cm^3}}} \right)^2, & \text{(Fixed by statistics \& dark matter.)}
\end{aligned}
\end{equation}
\end{widetext}
\begin{comment}
\begin{widetext}
\begin{multline}\label{eq:instscanspeed}
    \frac{df}{dt}~{\approx}~1.7~\frac{\mathrm{GHz}}{\mathrm{yr}} 
    %experimental
    \underbrace{ \left(\frac{\beta}{1 + \beta }\right)^2 
    \left(\frac{C}{0.4}\right)^2
    \left(\frac{V}{16 \times \SI{10.1}{\litre}}\right)^2
    \left(\frac{Q_{\text{L}}}{20{,}000}\right)}_{\rm \ref{sec:cavity}.~RESONATOR~DESIGN}
    \left(\frac{B}{\SI{9.1}{\tesla}}\right)^4
    \left(\frac{\SI{500}{\milli\kelvin}}{T_{\text{sys}}}\right)^2
    % statistics
    \left(\frac{3.5}{\text{SNR}}\right)^2 \\
    % dark matter
    \times \bigg(\frac{g_{\gamma}}{0.36}\bigg)^4
    \left(\frac{f}{\SI{2}{\giga\hertz}}\right)^2
    \left(\frac{\rho}{0.45\mathrm{~\mathrm{GeV/cm^3}}}\right)^2,
\end{multline}
\end{widetext}
\begin{multline}
    \frac{df}{dt}~{\approx}~543~\frac{\mathrm{MHz}}{\mathrm{yr}} \left(\frac{\beta}{1 + \beta }\right)^2 \left(\frac{g_{\gamma}}{0.36}\right)^4
    \left(\frac{f}{740~ \mathrm{MHz}}\right)^2
    \left(\frac{\rho}{0.45\mathrm{~\mathrm{GeV/cm^3}}}\right)^2
    \left(\frac{3.5}{\text{SNR}}\right)^2
    \left(\frac{B}{7.6~\mathrm{T}}\right)^4
    \left(\frac{V}{136~\mathrm{\ell}}\right)^2
    \left(\frac{Q_{\text{L}}}{30{,}000}\right) \\ \left(\frac{C}{0.4}\right)^2
    \left(\frac{\SI{200}{\milli\kelvin}}{T_{\text{sys}}}\right)^2,
\end{multline}
\end{comment}
\noindent where $\beta$ is the antenna coupling,
%${g_{\gamma}}$ is the dimensionless axion-photon coupling defined above
%$f$ is the resonant frequency,
$C$ is the mode's form factor (quantifying the mode's $\bm{E}$ overlap with the external $\bm{B}$),
$N$ is the number of resonant cavities, $V$ is each cavity's volume,
$Q_{\rm L}$ is the loaded quality factor of the mode (related to the unloaded quality factor $Q_0$ according to $Q_0 = (1+\beta) Q_L$),
$B$ is the average magnetic field in the cavity,
$T_{\text{sys}}$ is the system noise temperature, $\eta_c$ is the signal combining efficiency across all cavities,
$\text{SNR}$ is the target signal-to-noise ratio and
$\rho_\mathrm{DM}$ is the local DM density. Note that this equation is only valid if $Q_L\leq Q_a$, which is the case in this study.

In the following, we introduce the design of ADMX-EFR with an emphasis on the critical design parameters above, distinguishing between a \emph{threshold} design and an \emph{advanced} design. The \emph{threshold} design assumes conservative performance extrapolated from components used in the ADMX haloscope. The \emph{advanced} design is an estimate of what can be achieved using technologies that have been demonstrated but never employed by the main ADMX experiment.
%, such as quality factor enhancement from superconducting cavities.
% We believe the level of detail in our simulations, verified by measurements in current ADMX runs, goes well beyond what other sensitivity estimates presented in the field provide and is therefore a much more realistic estimate of what we will achieve. (probably better for a letter to the editor??)
Sec.~\ref{sec:magnet} outlines boundary conditions from magnet and cryogenic infrastructure. We detail our resonator design in Sec.~\ref{sec:cavity} and discuss readout electronics in Sec.~\ref{sec:electronics}. Particular attention is paid to coherent combining efficiency of signals from many resonators in Sec.~\ref{sec:powercomb}.
In Sec.~\ref{sec:sens_est} we use Eqn.~\eqref{eq:instscanspeed} and experimentally grounded assumptions of downtime to estimate the discovery potential of a 3-year experimental program. We conclude in Sec.~\ref{sec:conclusion}.

\section{\label{sec:magnet}Magnet and Cryostat}

%ADMX-EFR magnet parameters

\begin{figure*}
    \centering
    \includegraphics[width=\linewidth]{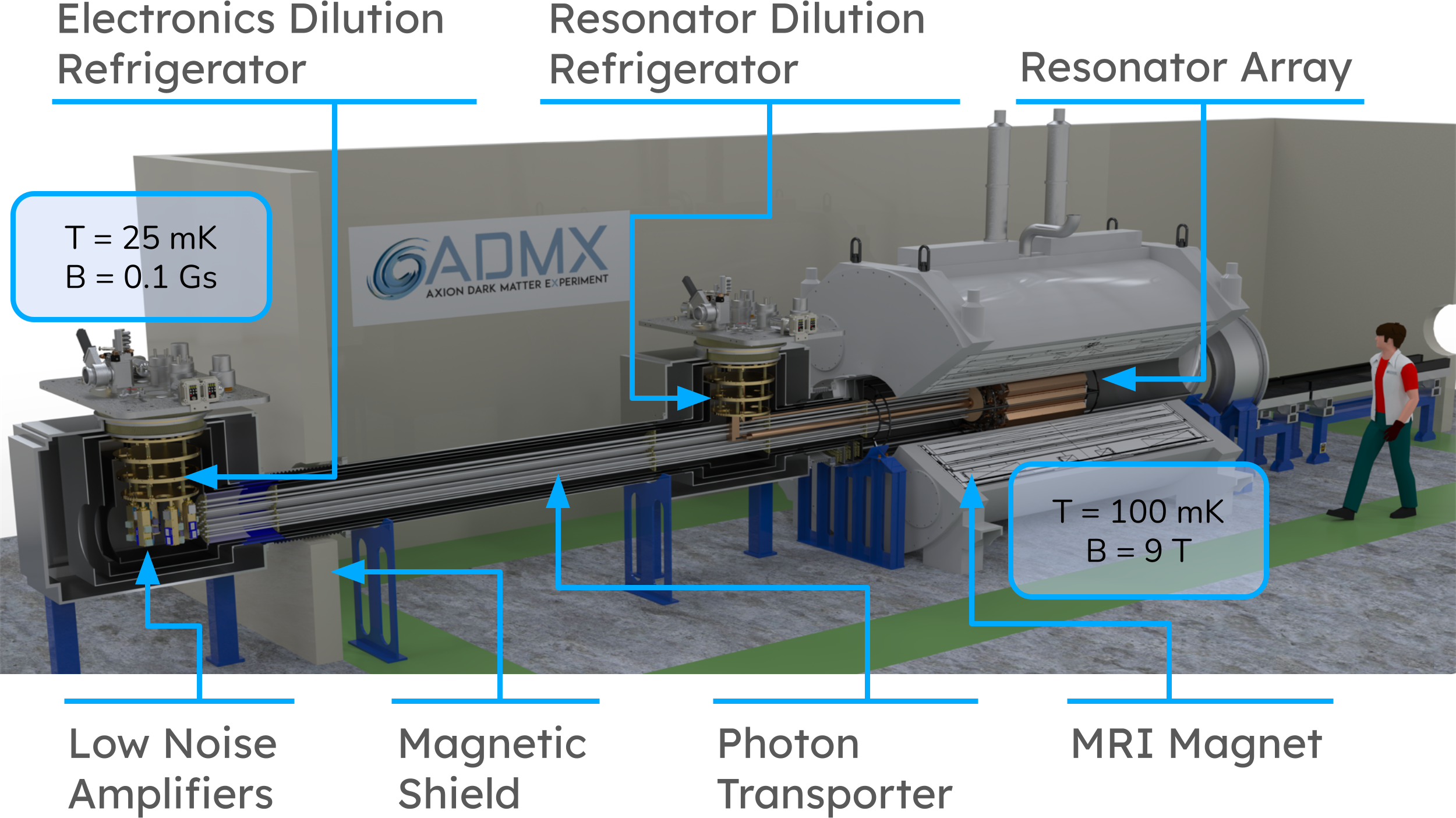}
    \caption{\label{fig:3d-overview} Schematic of the ADMX-EFR setup. The resonant cavities (right) are placed horizontally in a \SI{9.4}{\tesla} MRI magnet. They are cooled by a dedicated dilution refrigerator placed to the left of the magnet supported by additional cryocoolers placed on the opposite end of the magnet (not shown). The cavities are read out using separate low loss photon transport lines connecting them to a second dedicated dilution refrigerator in a shielded low-field region (left) where low noise quantum-limited amplifiers are housed. Nominal temperature and magnetic field strength in the electronics (left) and resonator (right) cryostats are quoted.}
\end{figure*}

%cryostat design and parameters
\noindent
For the purposes of this design study, we choose an MRI magnet with \SI{9.4}{\tesla} mean field and an \SI{800}{\milli\meter} bore diameter, matching the nominal specifications of a magnet currently located at Fermilab. This selection constrains the dimensions of the cavity array and custom-designed tandem dilution refrigerator cryostat. A rendering illustrating the magnet integration with other haloscope components is shown in Fig.~\ref{fig:3d-overview}.

\begin{figure}
    \centering
    \makebox[\linewidth][c]{\includegraphics[width=1.1\linewidth]{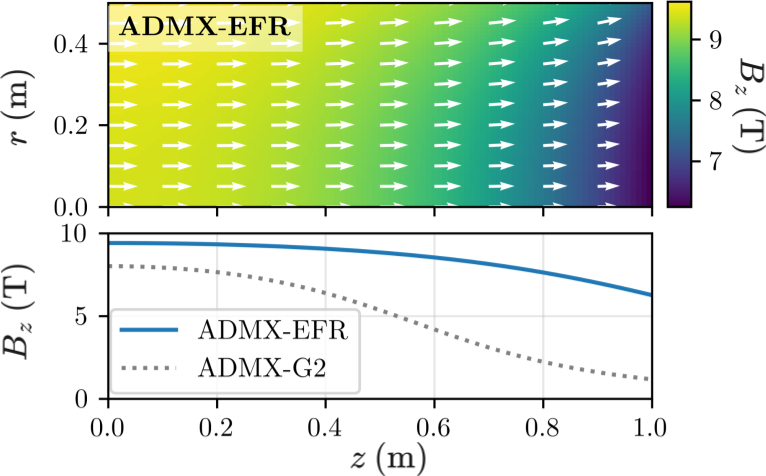}}
    \caption{\textbf{Top:} Magnetic field distribution in the ADMX-EFR magnet bore. Color indicates magnetic field strength in axial direction $B_z$ relevant for the axion search, arrows show magnetic field direction. \textbf{Bottom:}~Comparison of $B_z(z)$ along the central bore axis ($r=0$) between the ADMX-EFR magnet (blue solid) and the previously used ADMX-G2 magnet (gray dotted).} 
    \label{fig:bfield}
\end{figure}

The selected magnet was manufactured in 2003 by Magnex Scientific, under contract to GE Medical systems and was operated from 2023--2024 at University of Illinois Chicago as part of the first 9.4 Tesla whole-body MRI system ~\cite{1439885:uicmagnet}, before being relocated to Fermilab. Important design parameters compared to the magnet used in ADMX axion searches since 1993 are summarized in Tab.~\ref{tab:magnet}.
\begin{table}[b]
\begin{ruledtabular}
\begin{tabular}{lrr}
\textrm{}&
\textrm{ADMX}&
\textrm{ADMX-EFR}\\
& (1993-Present) & (this design) \\
\colrule
Nominal Field & \SI{7.6}{\tesla} & \SI{9.4}{\tesla} \\
Magnet Bore Diameter & \SI{0.53}{\metre} & \SI{0.80}{\metre} \\
Magnet Length & \SI{1.1}{\metre} & \SI{3.1}{\metre} \\
Cryostat Diameter & \SI{1.3}{\metre} & \SI{2.6}{\metre} \\
Stored Energy & \SI{16.5}{\mega\joule} & \SI{140}{\mega\joule} \\
\begin{tabular}{@{}l@{}}Max.\ $B_z$-Gradient \\ (over cavity region)\end{tabular}
%\footnote{Field gradient for the component along the bore axis over cavity region.} 
& \SI{100}{\milli\tesla\per\centi\metre} & \SI{25}{\milli\tesla\per\centi\metre} \\
[0.25em]
Weight & 6\,tons & 45\,tons \\
Helium Consumption & \SI{3}{\litre\per\hour} & \SI{0.4}{\litre\per\hour} \\
Current & \SI{204}{\ampere} & \SI{210}{\ampere} \\
Persistent Current & No & Yes \\
Orientation & Vertical & Horizontal \\
Manufacturer & Wang NMR & Magnex Scientific \\
Manufacture Date & 1993 & 2003 \\
\end{tabular}
\end{ruledtabular}
\caption{\label{tab:magnet}
Comparison of the ADMX-G2 magnet used in previous runs (e.g.~\cite{ADMX:2018gho}) and the new ADMX-EFR magnet.}

\end{table}
In order to estimate the impact of field inhomogeneity on axion searches performed in this magnet, a simplified simulation of the magnetic field is used, assuming uniform coils.
Fig.~\ref{fig:bfield} shows the simulated magnetic field as a function of radius $r$ and axial position $z$ where $(r,z) = (0,0)$ is the center of the magnet bore. The upper subplot shows the direction of the magnetic field and the magnitude of the $z$-component. The lower subplot compares the field along the central axis with the field of the ADMX-G2 magnet currently in use at the University of Washington. The Magnex magnet has a significantly higher magnetic field strength and a larger bore, enabling a greater axion sensitivity.

The magnet is shielded by a 136-ton 4-sided box made of C1006 low-carbon steel (dimensions \SI{5.4}{\meter} $\times$ \SI{12}{\meter}, with \SI{2.54}{\meter}-high walls). The magnet is centered on the symmetric shielding box in order to null reaction forces on the cold mass. The long ends of the shielding box have 500-\SI{}{\milli\meter} holes to allow the connection of the electronics dilution fridge with the magnet cryostat (See Fig.~\ref{fig:3d-overview}). Simulations show that the magnetic field in the region of the cold electronics is in the range 50--\SI{100}{\gauss}. As implemented in the current ADMX-G2 experiment design, individual two-layer Amumetal cans surround each cold electronics package, reducing the local field to below \SI{0.1}{\gauss}.

The magnet coils are immersed in an isolated bath of liquid helium, while the haloscope is cooled by two independent dilution refrigerators (DRs). The resonator DR is situated close to the magnet and cools the cavity array via a cold finger, while the electronics DR cools the field-sensitive RF amplifiers outside of the magnetic shielding.
The resonator DR is designed to minimize the distance between the cold radiation shields since the scan rate, Eqn.~\eqref{eq:instscanspeed}, scales strongly with available cryostat volume. The resulting available cold bore diameter is 676 \SI{}{\milli\metre}. Dedicated finite-element mechanical simulations show the cavity array support design withstands its \SI{650}{\kilo\gram} load without deflecting enough to produce a thermal short.

The electronics DR is located 5~m away from the cavity array, outside the magnetic shield, at the end of the \emph{photon transport} cables. The amplifiers are mounted on interchangeable frames dubbed \emph{parapets}. The electronics cryostat houses six identical parapets, each hosting three Josephson parametric amplifiers (JPAs), to serve the cavity array. To minimize downtime incurred by warm-ups and cool-downs resulting from reconfiguration or replacement of the cold electronics, each parapet is designed to be removed from the electronics cryostat without breaking vacuum or substantially heating the 18-cavity array. The vacuum system design is inspired by the Leiden cold-insertable probes \cite{Leiden}, while blind-mate RF connectors can join each parapet to the photon transport line. A separate test DR with a few parapet slots can be employed to test and calibrate the cold electronics prior to installation.
The implementation and implications of the photon transport, parapets, and other electronic components is discussed in Sec.~\ref{sec:electronics}.

\section{\label{sec:cavity}Resonator}
\begin{comment}
    
\emph{\color{blue} Gp /  }
\emph{\color{blue} Stefan gets updated data from Ben }
\emph{\color{blue} Ben / Nick / Tom  can chime in as they want}

design choices

results of Ben's simulations

mode crossings

quality factor considerations

\hline

\vspace{2em}
\end{comment}

\begin{figure}[h]
    \centering
    \includegraphics[width=\linewidth]{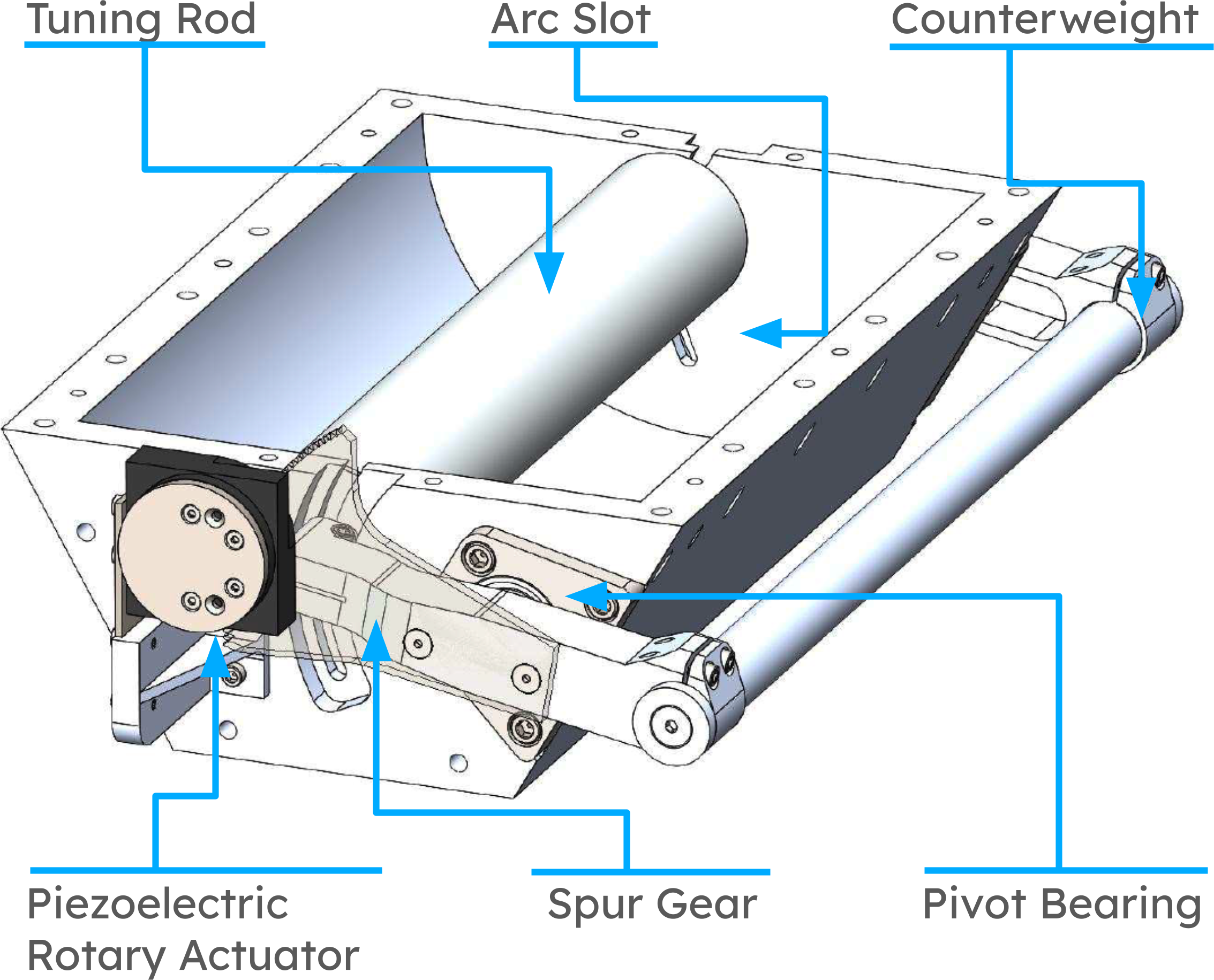}
    \caption{A single resonant cavity cell (a shortened test resonator is shown for illustrative purposes). As discussed in the text, the cavities are hexagonal in cross section and split longitudinally into two halves. They each contain a single counter-balanced tuning rod suspended on sapphire pins, with a rotation mechanism driven by a piezoelectric actuator. } 
    \label{fig:single_cavity}
\end{figure}

\begin{figure}[h]
    \centering
    \includegraphics[width=\linewidth]{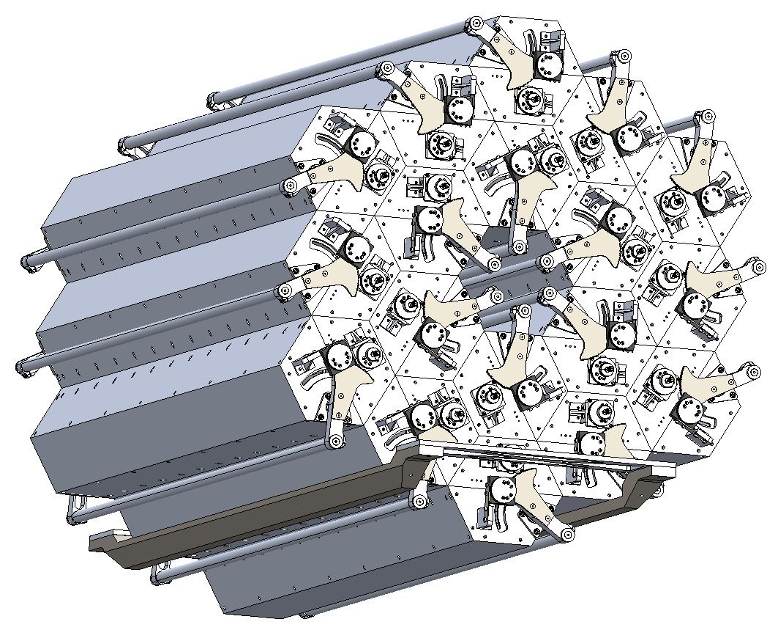}
    \caption{Rendering of the fully assembled 18-cavity array. All cavities are tuned and read out independently. The outer diameter of the assembly is chosen to make maximum use of the available space within the cryostat. The RF cabling for readout is not shown. }{\color{blue} } 
    \label{fig:resonator_array}
\end{figure}

\noindent
The ADMX-EFR design uses an array of 18 independent microwave cavity cells to resonantly enhance coherent axion to photon conversion (see Figs.~\ref{fig:single_cavity} and ~\ref{fig:resonator_array}). 

%Hence, cavity design parameters have a major impact on sensitivity as seen in eq.~\eqref{eq:instscanspeed}, comprising
%. Design parameters for a frequency (mass) scanning experiment include 
%Several of these parameters are inter-related and require trade-offs to optimize.
%Additional design considerations include the precision, thermalization and heat-loads of motion control elements.

Several cavity properties have a significant impact on the haloscope's ultimate sensitivity. The span of accessible axion masses is limited by the cavity tuning range $\Delta \nu$. The instantaneous scan rate depends on the cavity mode's form factor $C$, volume $V$, and loaded quality factor $Q_L$ (see Eqn.~\eqref{eq:instscanspeed}). The power draw and thermalization of the motion control elements used to tune the cavities also affect the rate at which the experiment can scan while maintaining a given $T_\mathrm{sys}$.

In this section, we focus on the design of a single resonator cell, as illustrated in Fig. \ref{fig:single_cavity}.  Coherent combination of the signals from multiple cavities is discussed in Sec.~\ref{sec:powercomb}.

%\subsubsection*{Individual Cavity}
\begin{comment}
    
    \centering
    \includegraphics[width=\linewidth]{figures/single_resonator.png}
    \caption{Conceptual Design of a single resonator cell shown as a cross section of the mechanical clamshell design, piezoactuator tuning mechanicsm using a central tuning rod held via slots at the ends of the cavity. The electric fields of the TM$_{010}$ mode obtained using a COMSOL simulation aligned with the cavity axis on the cross-sectional slice are shown. {\color{blue} [replace figure and update caption with rendering of more recent simulation / 1m mechanical design]} }
    \label{fig:cavity_mech_sim_fields}

\end{comment}

The individual unit cells of the ADMX-EFR cavity array are right circular cylinders measuring nominally \SI{1}{\metre} long and \SI{128}{\milli\metre} in diameter embedded in hexagonal prism bodies. Each cavity is split longitudinally into two ``clamshells.'' This design confers several advantages. The mode ideally induces electric currents that flow only vertically along the length
of the cylinder and not across the seam. Thus high $Q$ can in principle be maintained despite the nominally lossy seam (in short prototype cavities, $Q$ was improved by the addition of an indium seal). Additionally, this design provides easy access to the inside of the cavity for the deposition of high conductivity or superconducting films. 
%Finally, it allows the addition of a non-conducting gasket between clamshells to prevent eddy-current heating or hoop forces in the event of a magnet quench. 
% above omitted because DJT tried it and got bad Q
The clamshell cavity design has been demonstrated in a prototype called ADMX Sidecar~\cite{ADMX:2021mio}.

A conductive, movable tuning rod is mounted in the cavity which allows frequency tuning via adjustment of its radial position, similar to previous ADMX cavities~\cite{RevModPhys.75.777,ADMX:2020ote}. In ADMX-EFR, a set of \SI{32}{\milli\metre}-diameter tuning rods produces a resonance that can scan from $2.0$ to $3.1$\,GHz (\SI{12.1}{\litre} active volume), and a second set of \SI{54}{\milli\metre}-diameter rods is used to tune from $3.1$ to $3.9$\,GHz (\SI{10.1}{\litre} active volume). The radii of the cavity and the tuning rod are chosen to fill the available magnet bore with as many cavities as possible while covering the frequency range of interest. The tuning rod is suspended by sapphire pins through arc slots in the end caps, and a counterweight rod external to the cavity reduces motor load (see Fig.~\ref{fig:single_cavity}). An internal rod bearing design was also considered, similar to \cite{QUAX:2024fut}, but it was found that it would perturb the electromagnetic fields close to the end caps more strongly, leading to increased coupling to intruder modes discussed further below.

\begin{figure*}
    \centering
    \includegraphics[width=\linewidth]{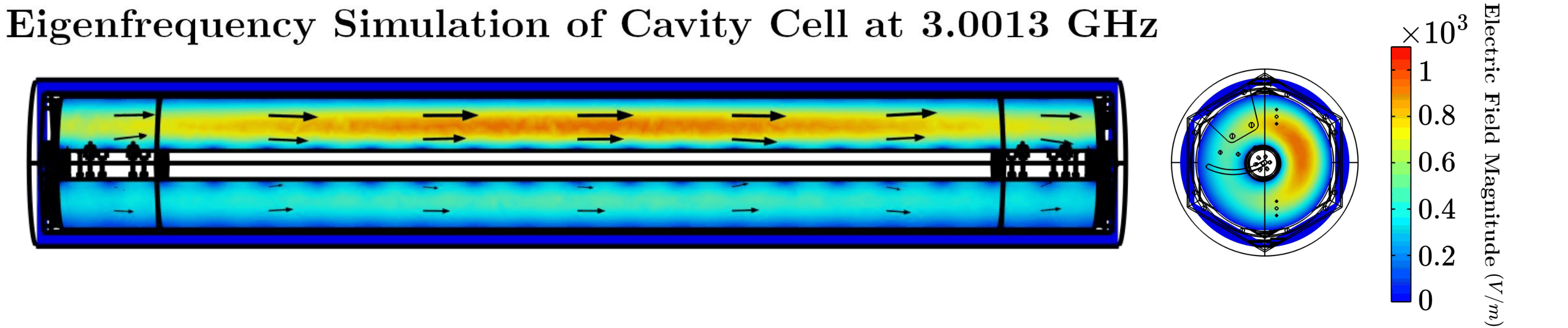}
    \caption{Electric field along the cylinder axis in the \SI{12.1}{\litre} cavity design tuned close to \SI{3}{\giga\hertz} (near the high end of its range), showing the TM$_{010}$-like mode with uniform electric field along the cavity's length. \textbf{Left:} Longitudinal cross section along the cavity axis. Black arrows show the electric field direction. \textbf{Right:} Transverse cross section.}
    %Conceptual Design of a single resonator cell shown as a cross section of the mechanical clamshell design, piezoactuator tuning mechanicsm using a central tuning rod held via slots at the ends of the cavity. The electric fields of the TM$_{010}$ mode obtained using a COMSOL simulation aligned with the cavity axis on the cross-sectional slice are shown. {\color{blue} [replace figure and update caption with rendering of more recent simulation / 1m mechanical design]} }
    \label{fig:cavity_mech_sim_fields}
\end{figure*}

To simulate resonator performance, we used the COMSOL RF module, CST Microwave Studio, and Ansys HFSS. Independent cross-checks were performed between simulation suites. Fig.~\ref{fig:cavity_mech_sim_fields} shows a rendering of a single cavity cell and the electric field within when tuned to a resonant frequency close to \SI{3}{\giga\hertz}. The electric field exhibits the uni-directional pattern typical of a TM$_{010}$-like mode, with large axion coupling form factor $C$. 

\begin{figure*}
    \centering
    \includegraphics[width=\linewidth]{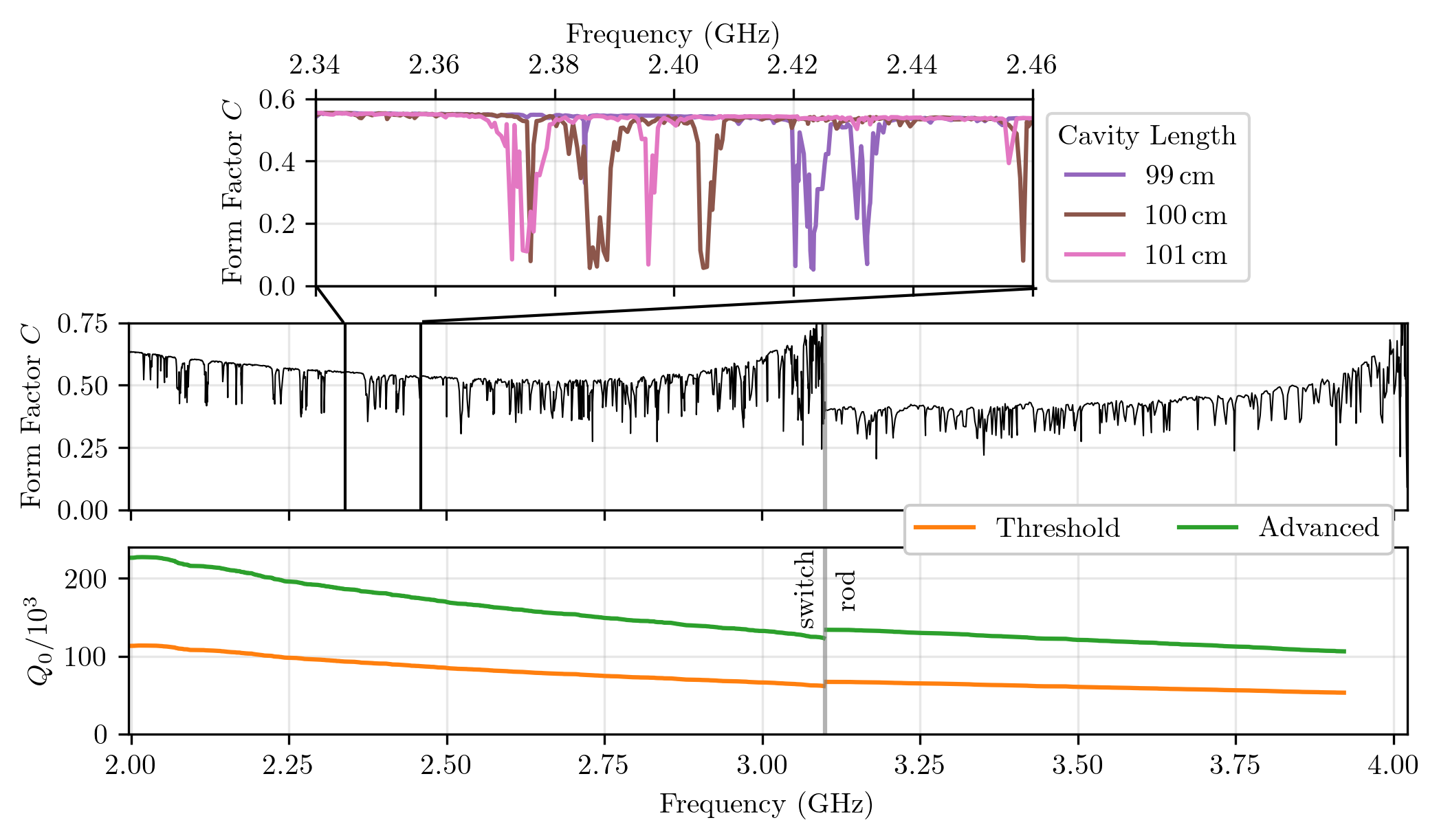}
    \caption{Simulated individual cavity design performance parameters. \textbf{Top:} Form factor $C$ of the TM$_{010}$-like mode for a \SI{120}{\mega\hertz} frequency range and three different cavity lengths. Drops in form factor are caused by mode crossings. The intruder modes can be detuned in frequency by changing the cavity length, allowing them to occur at different frequencies in a cavity array. \textbf{Middle:} Effective form factor of the full array over the full ADMX-EFR frequency range. The drops in $C$ are limited to $2/3$ of the full form factor, since three cavities of different lengths are used. The form factor is identical in both the \textit{threshold} and \textit{advanced} cases. The vertical gray bar here and below indicate the frequency at which the tuning rods must be exchanged to extend the cavity bandwidth. \textbf{Bottom:} Estimated unloaded quality factor for an entirely copper cavity (\emph{threshold}, orange) or a copper cavity with a superconducting tuning rod (\emph{advanced}, green) cavity. }
    \label{fig:cavity_performance_parameters}
\end{figure*}

The mode's form factor $C$ is calculated across the entire tuning range, accounting for the tuning rod exchange at $3.1$\,GHz. The cylindrical cavities also support TE-like ``intruder'' modes which do not tune strongly with rod position. When the TM$_{010}$-like mode's resonant frequency coincides with that of an intruder mode (known as a ``mode crossing''), they hybridize, leading to an effective reduction in $C$ and thus reduced sensitivity at intruder mode frequencies. Because intruder mode frequencies depend on cavity length, we can leverage the cavity-array design to mitigate this effect. By operating with 3 sets of 6 cavities, each of lengths \SI{990}{\milli\metre}, \SI{1005}{\milli\metre}, and \SI{1010}{\milli\metre}, there will always be at least one set of cavities unperturbed by a mode crossing (see Fig.~\ref{fig:cavity_performance_parameters}, top panel). The overall sensitivity of the experiment is set by the average $C$, which is shown in Fig.~\ref{fig:cavity_performance_parameters}, middle panel.
The form factor is identical in both the \textit{threshold} and \textit{advanced} cases.
Axial displacement of the tuning rod typically has a negligible impact on the form factor; however, it increases the coupling between the TM$_{010}$ and intruder modes, thus increasing the frequency ranges affected by mode crossings. Hence, all form factors presented conservatively assume maximal axial displacement of the tuning rod.

The unloaded quality factor $Q_0$ for a copper cavity cell at \SI{100}{\milli\kelvin} is shown in Fig.~\ref{fig:cavity_performance_parameters}, bottom panel. A nominal $Q_0$ is calculated by simulation, from which the \emph{threshold} and \emph{advanced} $Q_0$'s are derived. Simulations provide the geometric factor $G$ of the mode evaluated on each inner cavity surface, from which nominal $Q_0$ can be found via \cite{Braine:2022qal}
\begin{align}\label{eq:q}
    Q_0=\left(\sum_i\frac{R_{s,i}}{G_i}\right)^{-1}
\end{align}
for surface resistances $R_{s,i}$. At \SI{3}{\giga\hertz}, the $G$ of the cavity walls is $G_{\rm wall} \approx \SI{5e2}{\ohm}$, the tuning rod contributes $G_{\rm rod} \approx \SI{6e2}{\ohm}$ and the end caps $G_{\rm caps} \approx \SI{8e3}{\ohm}$. Eqn.~\ref{eq:q} shows that losses on the tuning rod and outer walls detract most from the quality factor. Using the anomalous skin depth of copper at cryogenic temperatures, $\delta_{\rm anom} \approx \SI{.2}{\micro\metre} \times (\SI{3}{\giga\hertz} / \nu)^{1/3}$~\cite{kittel:etde_7307886, Hagmann:1990ila}, one finds $Q_0\approx10^5$ at \SI{3}{\giga\hertz}. Tuning to a higher frequency makes the cavity electrically smaller, effectively increasing the surface-to-volume ratio for the mode, reducing the quality factor.
The slots that accommodate the tuning rods on the cavity end caps can lead to additional losses due to radiation. These must be minimized by optimizing the geometry of the slot. Simulations show that radiative loss through the slots is an order of magnitude less than surface resistive loss, and thus has a negligible effect on $Q_0$.
The \emph{threshold} $Q_0$ is reduced by a factor of 0.6 compared to the simulated value, a conservative scaling that accounts for any imperfections not accounted for (manufacturing tolerances, imperfect plating, environment-dependent leakage losses, etc.), consistent with the ratio between simulated and measured $Q_0$ in previous ADMX runs.
A range of recent work~\cite{Alesini:2019ajt,Golm:2021ooj,Kim:2022hmg,Posen:2022tbs,Cervantes:2022gtv,Braine:2022qal} shows the potential of superconducting RF cavities for axion searches, reducing cavity surface losses to near zero. Demonstrating the impact of this technology, the \emph{advanced} $Q_0$ assumes the use of superconducting tuning rods ($R_\mathrm{rod}=0$) and is thus double the \emph{threshold} value.

\section{\label{sec:electronics} Electronics} % RF?

\noindent
In this section, we describe the layout of the full cold electronics chain in EFR's two cryostats and noise temperature calibration procedures. Then, we discuss the custom-built flux-pumped JPAs that ADMX-EFR employs as first-stage amplifiers.

\subsection{Receiver Chains}
\begin{figure*}
    \centering
    \includegraphics[width=1.0\linewidth] {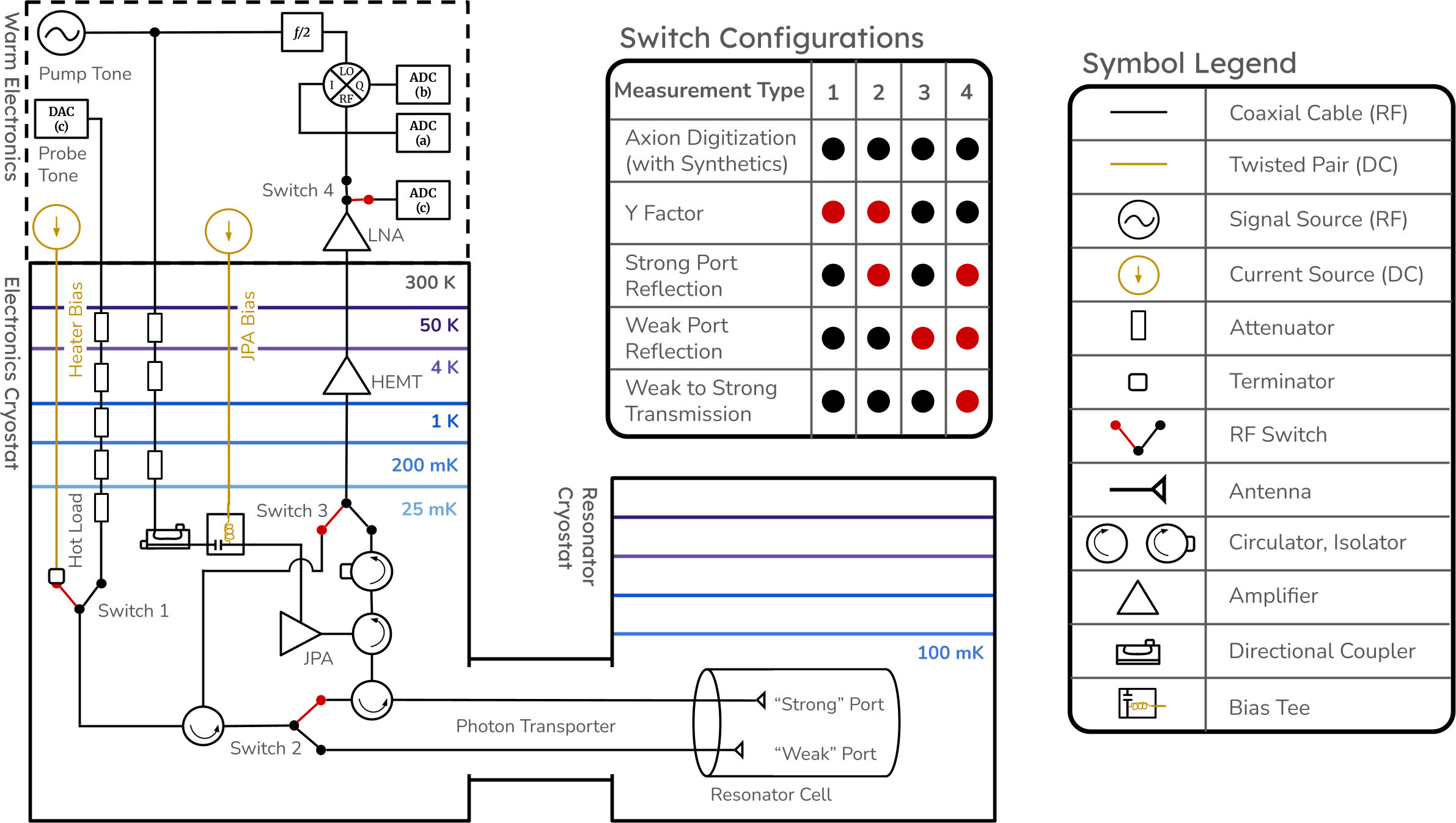}
    % {figures/ADMX_2_4_GHz_RFElectronics_20210504.pdf}
    \caption{
        Schematic of the receiver chain used to read out a single resonator cell in the ADMX EFR array, of 18 total. Each \emph{parapet} hosts three instances of the \SI{25}{\milli\kelvin} portion of this receiver chain. The chain can be configured using the four RF switches for axion signal digitization (normal operation), Y factor noise calibration, and resonator characterization. The colored dots in the ``switch configurations'' table specify which switch contacts are closed in each configuration. The ADCs \textbf{(a)} and \textbf{(b)} are used for digitizing signal power when in axion signal or Y factor configurations. The DAC-ADC pair \textbf{(c)} is used for resonator characterization measurements. When in the axion digitization configuration, the weakly coupled antenna port of the resonator may be used concurrently to inject synthetic axion signals for validation of the analysis pipeline.
        }
    \label{fig:rf_electronics}
\end{figure*}

\noindent
The cryogenic receiver chains in ADMX-EFR extract the low-power axion signal via the cavity antennas and amplify it into a signal detectable by room-temperature digitizing electronics. A schematic of the ADMX-EFR cryogenic electronics for three out of 18 cavities can be seen in Fig.~\ref{fig:rf_electronics}. As shown in the figure, each cavity is read out via its own receiver chain, with signals combined digitally during analysis (see Sec.~\ref{sec:powercomb}). The extraneous noise introduced by the receiver chain is characterized by the system noise temperature $T_\mathrm{sys}$, which impacts the experiment's sensitivity according to Eqn.~\eqref{eq:instscanspeed}. 
%The Friis equation for noise shows that $T_\mathrm{sys}$ is dominated by the noise at the input of the first-stage amplifier assuming its gain is greater than approximately \SI{20}{\decibel}.
\begin{comment}
Each amplifying stage in an electronics chain, with gain $G_i$ and input noise temperature $T_i$, contributes to $T_\mathrm{sys}$ according to the Friis formula:
\begin{align}
    T_\mathrm{sys} = T_1 + \frac{T_2}{G_1} + \frac{T_3}{G_1G_2} +\cdot\cdot\cdot+ \frac{T_n}{G_1\cdot\cdot\cdot G_n},
\end{align}
with elements numbered closest to farthest from the input.
\end{comment}

%The schematic in Fig.~\ref{fig:rf_electronics} shows the components mounted on a single parapet, as well as their connection to the cavities via the photon transport lines.

Here, we discuss each component in the receiver chain along the axion signal path. The signal is first coupled out of the cavity using a monopole antenna. The coupling $\beta$ to the cavity is adjusted by actuating the antenna, altering the insertion depth, analogous to previous ADMX runs (e.g. \cite{ADMX:2019uok}). 

Then, the signal propagates along the 5~m-long photon transport lines to the parapets. Because of their length, these lines must be very low-loss. A 10\% loss in signal power (\SI{-0.46}{dB}) requires a 20\% increase in integration time to maintain the same sensitivity in coupling $g_{\gamma}$. Transmission line losses can be reduced to a fraction of a dB by optimal deployment of superconducting cables and, in the high-field region,  specially selected (air core) normal conducting cables. The baseline choice for the photon transport lines is 0.5-inch AirCell coaxial cables with a predicted $\SI{0.4}{dB}$ loss over 5~m, calculated using the framework presented in \cite{CoaxAttenuation} and properties of the cable materials.

Two circulators lie between the terminus of the photon transport line and the JPA. Since they are on the input side of the JPA, circulators are one of the leading sources of initial loss, and therefore noise, in axion haloscope experiments. For this study, Low Noise Factory components LNF-CIC2\_2.8A \cite{LNF:2-2.8} and LNF-CIC2.8\_4.3A \cite{LNF:2.8-4.3} are selected, covering the 2.0$-$\SI{2.8}{\giga\hertz} and 2.8$-$\SI{4.0}{\giga\hertz} ranges respectively. The circulator bandwidths are such that one parapet exchange does not coincide with the tuning rod exchange discussed above. The JPA tuning ranges can be engineered to coincide with the circulator bandwidths. The circulators each have an extremely low \SI{0.2}{\decibel} insertion loss and their isolation is at least \SI{18}{\decibel}.
%To address this issue, custom circulators have been developed in collaboration with Quinstar. The circulators provided by Quinstar are double-junction devices which could \td{have they been tested?} achieve $<\SI{0.3}{dB}$ insertion loss and $>\SI{40}{dB}$ of isolation over a % \SI{1}{GHz} $\sim\SI{0.5}{GHz}$ frequency range. This is a marked improvement over past devices which typically had  $\sim\SI{0.3}{dB}$ insertion loss per single junction, and $\sim\SI{20}{dB}$ isolation. 
An isolation greater than the JPA gain is important to prevent amplified signals from reflecting back to the cavity resonators, producing spurious RF cable modes.
%With a 0.5-GHz bandwidth, the circulators are the limiting components in this respect, requiring three exchanges. Downtime is discussed in detail in Sec.~\ref{sec:sens_est}.

The signal is then amplified at the JPA (discussed below in Sec.~\ref{sec:jpa}) and travels through a multi-junction isolator. At the \SI{4}{\kelvin} stage, it encounters the second-stage amplifier (the LNF-LNC1.5-6A), which covers the entire EFR range with a gain of \SI{28}{dB} and noise temperature less than $\SI{1.8}{K}$.

Finally, the signal passes to the room-temperature receiver chain to be processed and digitized for analysis. ADMX-EFR uses a software-defined radio (SDR) platform, the Ettus Research USRP X300. Each X300 provides two ADC channels, two DAC channels, and an integrated field-programmable gate array (FPGA). Custom FPGA firmware, together with software running on a dedicated server, supplies the digitization, signal processing, resonance monitoring, and signal injection capabilities that previously required a separate digitizer, vector network analyzer (VNA), and arbitrary waveform generator (AWG). Each X300 serves two cavities, thus only nine units can read out the full array.

In the room-temperature receiver, the signal is further amplified and passed through a set of analog filters, conditioning it to match the input voltage range of the digitizer. The signal is then digitized and digitally downconverted to the band of interest around the cavity resonance. This process retains only the frequency range in which an axion signal could appear, reducing the data volume and making simultaneous readout of all 18 resonators tractable. The resulting time series data is then written to disk.

The X300 can also monitor the steady-state response of each cavity with a VNA-style measurement, in which a tone generated on a DAC channel is swept across a chosen bandwidth and the cavity response is recorded.
%(see \textbf{c} components in Fig.~\ref{fig:rf_electronics})
In addition to providing resonance parameters necessary for axion search analysis, this measurement is used to track each cavity's resonant frequency and to drive the corresponding tuning rod, locking all 18 cavities to a common resonant frequency. During development of the receiver, a VNA implemented on the X300 was shown to reproduce the resonance parameters of an RF resonator measured with a commercial VNA.

A set of RF switches in each parapet allows for the full suite of characterization measurements to be carried out rapidly. RF switch configurations required to perform each type of measurement are given in the ``switch configurations'' table in Fig.~\ref{fig:rf_electronics}, where the dots' colors represent which contact should be closed for each of the four switches in the diagram.
During digitization, axion signal power is coupled out of the resonator at the strongly coupled port and passed through several amplifiers before being downmixed and digitized by ADCs \textbf{(a)} and \textbf{(b)}. Concurrently, for calibration and analysis validation, synthetic axion signals generated by DAC \textbf{(c)} may be coupled into the resonator via the secondary weakly coupled port.
Resonator characterization is accomplished by performing VNA-style reflection transmission measurements with tone originating at DAC \textbf{(c)} and digitized at ADC \textbf{(c)}. During typical data taking, resonator parameters are determined via a transmission from the weakly coupled to the strongly coupled port, which requires no cryogenic RF switch actuations, alleviating heat load on the DR.

At regular intervals, JPA noise temperature is measured using the Y-factor method. The hot load, a heated terminator on the input to the resonator, is set to various elevated temperatures and the resulting output power is measured at ADCs \textbf{(a)} and \textbf{(b)}. A tunnel-junction noise source, where the injected noise is controlled by a voltage and does not require a long thermal equilibration time between measurements, would allow for more rapid Y-factor measurements \cite{NoiseSource}. Since the cavities are cooled by a second DR, heating them separately from the cold electronics is feasible in our setup and can provide an absolute calibration of either noise source.

The total system noise temperature of the experiment was simulated across its bandwidth using the \pkg{scikit-rf} Python package \cite{skrf}. The simulation encompassed every component and transmission line along the axion's signal path from the strongly coupled port of the resonator to the output of the high-electron-mobility transistor (HEMT) amplifier, after which the signal is sufficiently amplified that loss has a negligible effect on SNR.

\subsection{Flux-Pumped Josephson Parametric Amplifiers}\label{sec:jpa}
\noindent
A flux-pumped JPA \cite{yamamoto2008} is a two port device that includes a  {\it signal port} for the incoming signal and outgoing amplified signal and a {\it flux pump port} for the flux pump tone and DC flux bias.
Signals incident on the signal port are coupled to an LC resonant circuit, where the inductor is realized as a series of superconducting quantum interference device (SQUID) loops whose inductance is controlled by the combination of a DC flux bias (tuning the linear resonant frequency) and an oscillating, small-signal flux pump tone (supplying the power necessary for amplification).
Parametric gain is achieved by applying a pump tone with a frequency equal to twice that of the input signal.

Flux-pumped devices offer some key advantages over the current-pumped JPAs employed in earlier configurations of the ADMX-G2 experiment.
In both device types, the JPA is biased to achieve the optimal operational mode by tuning three parameters: magnetic flux bias $\Phi_\mathrm{bias}$, pump power $P_\mathrm{pump}$, and pump frequency $\nu_\mathrm{pump}$~\cite{JPAreview}.
In the current-pumped case, increasing $P_\mathrm{pump}$ shifts the resonant frequency of the device, requiring a new shift in $\Phi_\mathrm{bias}$. Consequently, the tuning procedure involves alternating steps of $P_\mathrm{pump}$ and $\Phi_\mathrm{bias}$ to obtain optimal gain. For a flux-pumped JPA, changing $P_\mathrm{pump}$ does not shift the resonant frequency, so one may tune $P_\mathrm{pump}$ and $\Phi_\mathrm{bias}$ separately, greatly reducing operational complexity. Additionally, for a current-pumped JPA, the optimal $\nu_\mathrm{pump}$ is the center frequency of the desired gain curve, i.e. the axion signal frequency. This allows unwanted leakage from the pump tone into the signal band if the pump is not sufficiently detuned from its optimum. For a flux-pumped device, the $\nu_\mathrm{pump}$ optimum is twice the gain frequency, so there is no risk of such leakage.

\begin{figure}
    \centering
    \includegraphics[width=0.9\linewidth]{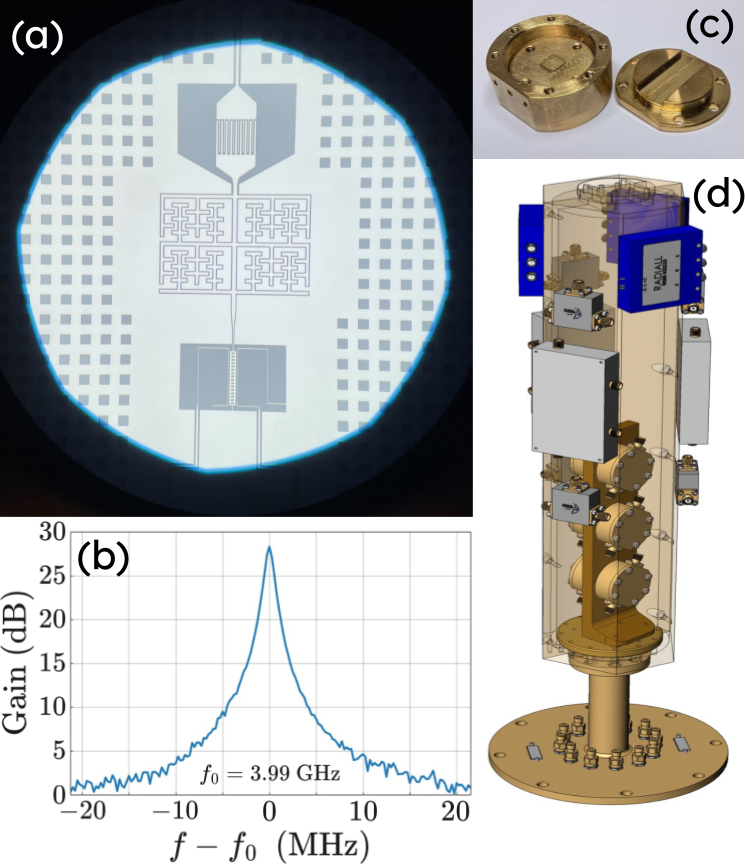}
    \caption{
        (a) Optical microscope image of a fabricated 3-\SI{4}{GHz} prototype flux-pumped JPA,
        (b) a representative gain curve,
        (c) custom enclosures, modified to move box modes away from relevant frequencies, and
        (d) mechanical design of one {\sl parapet}, which accommodates 3 cold electronics channels.
        The pictured configuration includes the JPAs, circulators, and RF switches for the noise source calibration.
    }
    \label{fig:parapet}
\end{figure}

The first flux-pumped JPAs for ADMX, fabricated at Washington University in St. Louis, were designed for amplification around \SI{1}{\giga\hertz}. In this regime, achieving appreciable parametric gain is more challenging than at higher frequencies.
The large planar capacitor and long traces in the device can add a significant parasitic linear inductance to the circuit. While multilayer capacitors reduce this effect, they suffer from higher loss tangents. We employ a fractal, interdigitated planar capacitor design allowing us to benefit from the low-loss of planar geometry while minimizing parasitic inductance. The structure achieves proximity cancellation of the magnetic fields generated by opposing currents, reducing the magnetic energy density and hence the inductance.
To achieve the required tuning range we developed a new type of differential pumping scheme, allowing us to include a large number of squid-loops (16) with pure flux coupling and reduced (unwanted) common mode coupling. This scheme makes use of stitched ground planes using wire-bonds rather than using multi-layer connections. 
These features can be seen in the image of the device shown in Fig.~\ref{fig:parapet}, panel~(a).
%\td{detail required?}
%\td{it's unclear to me which design elements translate to the EFR design}

The combination of the above design features provides a very high ratio of nonlinear to linear inductance, allowing us to achieve the design specifications: greater than $\SI{25}{dB}$ gain
%, $>\SI{15}{dB}$ signal-to-noise-ratio improvement (SNRI) \td{convert to $T_N$?} 
over a broad \SI{400}{\mega\hertz} tuning range from 1.1 to \SI{1.5}{\giga\hertz} and at least \SI{15}{\decibel} of compression for out-of-phase signals.
Thus, in a more challenging low-frequency design aimed at future improvements to the ADMX-G2 experiment, we developed techniques to meet the requirements for the ADMX-EFR JPAs. The nominal gain and added noise characteristics of these devices are quoted in Tab.~\ref{tab:parameters}.

We have since used these new design features to produce the 2--\SI{3}{GHz} and 3--\SI{4}{GHz} JPAs, the first of which have been characterized. A sample gain curve for a single pump frequency is shown in Fig.~\ref{fig:parapet}, panel~(b).
At higher frequencies the capacitor is smaller, decreasing the linear inductance and allowing for more tunability.

A custom gold-plated oxygen-free high conductivity copper housing was designed for these devices, pictured in Fig.~\ref{fig:parapet}, panel~(c). Two SMA panel connectors with center pins connected to the coplanar waveguide lines of a launch board provide the flux pump and signal port connections. Extrusions on the enclosure cap push box modes above the JPA's operational frequency. As discussed in Sec.~\ref{sec:magnet}, the JPAs are mounted on frames called parapets, with each parapet housing the RF components necessary for three receiver chains each (see panel~(d) in Fig.~\ref{fig:parapet}).

%A 2-port enclosure was developed for the ADMX-EFR JPAs by modifying the former design employed in the UCB JPAs used in the ADMX-G2 experiment (see Fig.~\ref{fig:parapet}, panel b). The enclosures are made of gold-plated OFHC copper components, including the main enclosure body with two SMA panel connectors with center pins connected to the CPW lines of a launch board to provide the flux pump and signal port connections. The launch board is fabricated from a Rogers substrate which has proven to exhibit good properties at cryogenic temperatures.  A separate copper pedestal spool includes the bias coil (with $\sim 200$ turns of NbTi wire wound around the spool) and a square pad for mounting the JPA silicon device.  A hole in the launch board (between the two CPW traces) allows the JPA device to be mounted level with the CPW lines, providing good impedance matching, device thermalization and strain relief for the wirebonds.     
%Recent optimations of the enclosure include
% pillars on 
%extrusions on the cap of the enclosure to push box modes to higher frequencies, away from the JPA's operational frequency. \td{I think discussion of the enclosures could be reduced}

\section{\label{sec:powercomb} Signal Combination}

\noindent
Signals from each resonator cell in the ADMX-EFR array are combined to produce a limit on axion-photon coupling.
Such combining schemes have been proposed and demonstrated in other experiments~\cite{RevModPhys.75.777, JEONG201833,Adair:2022rtw}.
%Such coherent power combination has been proposed a while ago in the community~\cite{RevModPhys.75.777, JEONG201833} and recently been demonstrated in a small-scale axion search~\cite{Adair:2022rtw}.

A direct approach to combining the signals from each of $N$ cells is passive combination onto a single RF line using a Wilkinson microwave combiner before amplification. While employing a single receiver chain substantially reduces operational complexity and data volume, there are consequential disadvantages. 
First, malfunctioning cells (with, for example, jammed rod actuators or nonfunctional amplifiers) cannot be excised, resulting in an imbalance of voltage across the inputs and thus dissipation in the combiner. If $M$ cells malfunction and no longer contribute signal, the combined power $P_N$ from the array is scaled by $((N-M)/N)^2$ \cite{Wilkinson1960}. For concreteness, a passively combined array with two malfunctioning cells ($M=2$) has 62\% of the instantaneous scan rate of a fully functional one. Further, the relative phases of signals from each cell cannot be adjusted to account for phase differences in the system without additional physical components that contribute to operational complexity themselves, and increase insertion loss and thus noise. Phase-coherent combination is essential for the resonator array haloscope and is discussed further below. To counteract these problems, each resonator in ADMX-EFR is instead read out independently on its own electronics chain, and the signals are combined digitally. In this configuration, for $M$ malfunctioning cells, $P_N$ scales as $(N-M)/N$ and thus $M=2$ entails only a 79\% scaling of instantaneous scan rate, illustrating that an independently read out array is more resilient to malfunction than a passively combined array.

For an independently read out array to achieve optimal sensitivity, the phases from each combined signal must be aligned. There are many potential sources of relative phase shift throughout the signal's path from the axion field itself to the digitizers. First, consider the spatial coherence of the axion field. Since the axions in the DM models we consider are non-relativistic, their de Broglie wavelengths are three orders of magnitude larger than the converted photons' wavelengths, $\lambda_{\rm dB} = h / (v_a m_a) \approx 10^{3} \lambda_\gamma \approx \SI{100}{\metre}$. Thus, the axion field (and thus signal) is phase coherent across the physical extent of the resonator array. The next consideration is the coherence time of the axion field, $\tau_a=Q_a/\nu_a$, which for ADMX-EFR is at worst $\tau_a=10^6/\SI{4}{\giga\hertz}=\SI{250}{\micro\second}$. This emphasizes the need for simultaneous readout of every resonator cell in the array. Signals collected from different cells at the same frequency but different times cannot be advantageously combined.

With phase delays intrinsic to the axion field handled, we turn now to those delays created by the experimental apparatus. The path length differences in the electronics chains of the cells create a fixed relative phase $\Delta\phi^{\rm fixed}_j$ between cell $j\in[1,N]$ and a common reference. Differences in the tuning of the JPAs can produce phase changes which change each time the JPA is biased, $\Delta\phi^{\rm JPA}_j$. Finally, if the resonator is detuned compared to a chosen target frequency, a phase delay is introduced which changes each time the tuning rod is moved $\Delta\phi^{\rm res}_j$. 

The goal of digital phase correction is to introduce, during analysis, new phase shifts $\Delta\phi^{\rm corr}_j$ in the signals from each cell which cancel out the phase delays introduced above, allowing for optimal signal combination. The non-cavity phase shifts $\Delta\phi^{\rm fixed}_j$ and $\Delta\phi^{\rm JPA}_j$ can be measured using an off-resonance tone from a single source, split and injected in-phase into the receiver chains of each resonator. The resonator phase shift $\Delta\phi^{\rm res}_j$ can be calculated analytically from its detuning from the target frequency $\Delta \nu_j$ and its quality factor $Q_j$ (parameters already calculated at each tuning step) using the standard Lorentzian resonator model. Note that this phase correction does not mend the inherent signal \textit{magnitude} loss incurred by cavity detuning from the axion frequency.

\begin{comment}
Total voltage-combined signal power at frequency $f$ can be expressed as

\begin{align}\label{eq:voltagesum}
    P_N(f) &= \frac{1}{Z}\left | \sum^N_{k=0} \tilde{G}_{k,0}(f)\tilde{V}_k(f) \right |^2,
\end{align}
\noindent
where $\tilde{V}_k(f)$ is the voltage signal from resonator cell $k$ subject to phase shifts from the apparatus and the compensating phase shift introduced digitally:
\begin{align}    
    \tilde{G}_{k,0}(f)&=e^{-i\Delta\phi^{\rm corr}_k}.
\end{align}    
\end{comment}

To elucidate the quantitative requirements on phase coherence, let us describe the signal combination formally. Our goal is to determine the dependence of the signal to noise ratio (SNR), a quantifier of haloscope sensitivity, on the $\Delta\phi_j^\mathrm{corr}$ error, assumed to be normally distributed about zero with variance $\sigma_\phi^2$.

We express the complex voltage of the axion signal at a given frequency $\nu_a$ from resonator cell $j\in[1,N]$ as
\begin{align}
    s_j &= ae^{i\phi_j},
\end{align}
with fixed, common amplitude $a$ and a cell-dependent phase offset $\phi_j$. An extension to distinct or time-dependent $a$ is left to a later work. In this analysis, we assume digital phase correction has been applied, and $\phi_j$ is the residual phase relative to the common true axion phase. As such, $\phi_j$ is drawn from a normal distribution with zero mean and variance $\sigma_\phi^2$. Note that since $\phi_j\in[0,2\pi)$, its distribution becomes uniform as $\sigma_\phi\to\infty$. We use $\langle\cdot\rangle$ to denote a mean over samples.

We begin by finding the covariance matrix for the signals from each cell. Its diagonal elements are given by
\begin{align}
    S_{jj} = \langle s_j s_j^*\rangle = a^2
\end{align}
and its off-diagonal elements are
\begin{align}
    S_{ij} = \langle s_i s_j^*\rangle = a^2 \langle e^{i\phi_i} \rangle \langle e^{-i\phi_j} \rangle.
\end{align}
The mean can be split among the phase terms because they are independent for $i\neq j$. Completing the Gaussian probability-weighted integral yields $\langle e^{\pm i\phi_i} \rangle = e^{-\sigma_\phi^2/2}$. We define 
\begin{align}
    \gamma\equiv e^{-\sigma_\phi^2}
\end{align}
and conclude that for $i\neq j$, 
\begin{align}
    S_{ij} = a^2\gamma.
\end{align}
The parameter $\gamma\in[0,1]$ quantifies the degree to which the signals are in phase, on average (across $\phi_j$ draws), for a given $\sigma_\phi$. 

With $S_{ij}$ completely determined, we may find the $N$ collective eigenmodes of the array. The common $k=1$ mode has unit eigenvector $\bm{u_1}=\frac{1}{\sqrt N}\begin{pmatrix} 1 & 1 & ... & 1\end{pmatrix}$. The $N-1$ other eigenvectors, for $k>1$, can be any orthonormal completion of the basis and have identical eigenvalues. The eigenvalues are denoted $\lambda_k$.
The eigenmodes are, in terms of the individual cell voltages $\bm{s}=\begin{pmatrix}s_1 & ... & s_N\end{pmatrix}$,
\begin{align}
    m_k = \bm{u^\dagger_k}\bm{s}.
\end{align}

We are now prepared to calculate the SNR that each mode provides. The mean measured axion signal power in mode $k$ is given by
\begin{align}
    \langle P_k \rangle &=\langle | m_{k}|^2\rangle \\
    &= \bm{u^\dagger_k} \langle \bm{s} \bm{s^\dagger} \rangle \bm{u_k} \\
    &= \bm{u^\dagger_k} S \bm{u_k} \\
    &= \lambda_k.
\end{align}
The noise variance in a single resonator cell is the usual $\sigma_n^2=k_BT_\mathrm{sys}\Delta\nu$ for measurement bandwidth $\Delta\nu$.
\begin{comment}
The noise in a single resonator cell $j$ at a particular time sample $t\in[1,M]$ can be expressed as
\begin{align}
    n_{j,t} = x_{j,t} + iy_{j,t},
\end{align}
with both $x_{j,t}$ and $y_{j,t}$ drawn from $\mathcal{N}(0,\sigma_n)$.
Note that for measurement bandwidth $\Delta\nu$ and integration time $\tau$, we have $M=\Delta\nu\tau$ and $\sigma_n^2=k_BT_\mathrm{sys}\Delta\nu$.
$\operatorname{Var}(P_k)$ 
\end{comment}
The noise is basis-independent, so the noise in a given mode is the same as the noise in a single cell.
The pre-integration SNR of mode $k$ is thus
\begin{align}
    \mathrm{SNR}_k = \frac{\langle P_k \rangle}{\sigma_n^2}=\frac{\lambda_k}{\sigma_n^2}.
\end{align}
For the common mode, $k=1$, % set a = \sigma_n = 1 ? I like that it's complete...
\begin{align}
    \mathrm{SNR}_{k=1}(\gamma) = \frac{a^2}{\sigma_n^2} (1+(N-1)\gamma),
\end{align}
writing SNR as an explicit function of $\gamma$. Note that $\mathrm{SNR}_{k=1}(1)=Na^2/\sigma_n^2$ and $\mathrm{SNR}_{k=1}(0)=a^2/\sigma_n^2$.
For $k>1$,
\begin{align}
    \mathrm{SNR}_{k>1}(\gamma) = \frac{a^2}{\sigma_n^2} (1-\gamma).
\end{align}
Examining the same extrema, $\mathrm{SNR}_{k>1}(1)=0$ and $\mathrm{SNR}_{k>1}(0)=a^2/\sigma_n^2$. In all, when $\gamma$ is large, the axion signal occupies the common mode alone, and when $\gamma$ is small, the axion signal is distributed evenly across modes. 

As an aside, at low $\gamma$, SNR can be improved by measuring total signal power from all cells rather than the common mode power. Signal power adds linearly while noise adds in quadrature, for a total power-combined SNR of
\begin{align}
    \mathrm{SNR}_\mathrm{tot} = \frac{Na^2}{\sqrt N\sigma_n^2} = \sqrt N\frac{a^2}{\sigma_n^2}.
\end{align}
We find that for $N=18$ cells, $\mathrm{SNR}_\mathrm{tot}>\mathrm{SNR}_{k=1}$ for $\gamma<0.191$, or $\sigma_\phi > 74^\circ$. Beyond that phase correction error, the cells can be considered effectively independent. Because of the gains in SNR, we instead aim to maximize $\gamma$ and read out common mode power. Below, SNR refers to SNR$_{k=1}$.

We quantify the impact of phase correction error on the instantaneous scan rate via the \emph{combination efficiency} $\eta_c$, the ratio of measured SNR($\gamma$) to ideal SNR$(1)$:
\begin{align}
    \eta_c=\frac{\mathrm{SNR}(\gamma)}{\mathrm{SNR}(1)}=\frac{1+(N-1)\gamma}{N}.
\end{align}
Note that $\eta_c\in[1/N,1]$ --- SNR is bounded from below by that of one cell.
The scan rate depends on SNR($\gamma$) as
\begin{align}
    \frac{\mathrm{d}\nu}{\mathrm{d}t} \propto \mathrm{SNR}(\gamma)^2&=\left(\frac{\mathrm{SNR}(1)}{\mathrm{SNR}(1)}\mathrm{SNR}(\gamma)\right )^2\\&=\left (\eta_c\mathrm{SNR}(1)\right)^2.
\end{align}
In fact, the SNR that appears in Eqn.~\ref{eq:instscanspeed} is SNR(1), since the benchmark scan rate used in the prefactor assumes no SNR loss due to phase incoherence. In other words, perfect phase coherence ($\gamma=1$) allows the array to behave as a single resonator with volume $NV_\mathrm{cell}$. 
Also note that the above analysis considered pre-integration SNR for simplicity, while it is the \emph{post-}integration SNR that appears in the scan rate. The result holds true, however, because the SNR increase from integration is independent of $\gamma$. 
In this section, we quantified the scan rate degradation compared to the ideal $\gamma=1$ case with the parameter $\eta_c$, which is derived from the digital phase correction error $\sigma_\phi$.

\section{\label{sec:sens_est}Discovery Potential}

\begin{figure}
    \centering
    \includegraphics[width=\linewidth]{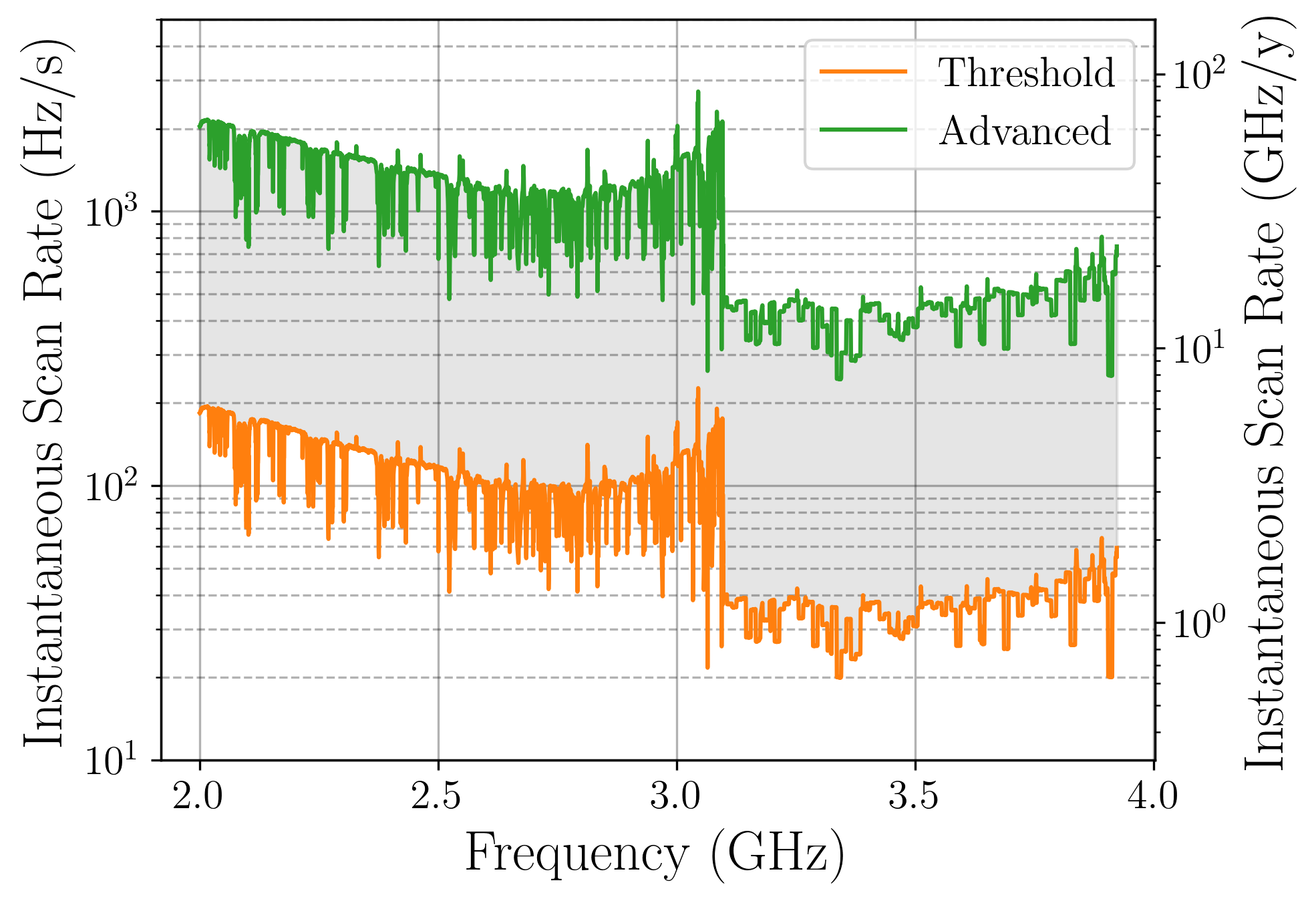}
    \caption{\label{fig:scanrate} Instantaneous scan rate estimated for ADMX-EFR with DFSZ sensitivity to axions with local density $\rho_\mathrm{DM}=\SI{0.45}{\giga\electronvolt\per\centi\meter\cubed}$ at SNR$=3.5$. 
    The \emph{threshold} and \emph{advanced} set of design parameters are detailed in Tab.~\ref{tab:parameters}.
    Grid lines align with the left-hand axis only.}
\end{figure}

\noindent
Tab.~\ref{tab:parameters} summarizes critical design parameters discussed in each previous section.
\begin{table}[b]%The best place to locate the table environment is directly after its first reference in text
\begin{threeparttable}
\begin{ruledtabular}
\begin{tabular}{lrr}
\textrm{Design Parameter}&
\textrm{Threshold}&
\textrm{Advanced}\\
\colrule
\textbf{Magnet (Sec.~\ref{sec:magnet})} & & \\
Mean B-Field $B_0$ & \SI{9.4}{\tesla} & \SI{9.4}{\tesla}  \\
\textbf{Resonator (Sec.~\ref{sec:cavity})} & & \\
Tuning Range & \multicolumn{2}{c}{$2.0-$\SI{4.0}{\giga\hertz} }  \\
Number of Resonators $N$ & $18$ & $18$ \\
Resonator Volume $V$ & \multicolumn{2}{c}{\SI{12.1}{\litre} ($2-$\SI{3}{\giga\hertz})}  \\
 & \multicolumn{2}{c}{\SI{10.1}{\litre} ($3-$\SI{4}{\giga\hertz})} \\
Quality Factor $Q_0$\tnote{$\dagger$} & $85,300$ & $170,000$ \\
Form Factor $C_{\rm TM010}$\tnote{$\dagger$} & $0.50$ & $0.50$ \\
Cavity Physical Temperature & \SI{120}{\milli\kelvin} & \SI{120}{\milli\kelvin} \\
\textbf{Electronics (Sec.~\ref{sec:electronics})} & & \\
Insertion Loss from Cavity to JPA\tnote{$\dagger$}& \SI{1.5}{\decibel} & \SI{1.5}{\decibel} \\
JPA Peak Gain & \SI{15}{\decibel} & \SI{21}{\decibel} \\
%JPA / Circulator Bandwidth& \SI{0.5}{\giga\hertz} & \SI{1}{\giga\hertz} \\
JPA Added Noise Photons & 2 & 1 \\ % & \SI{383}{\milli\kelvin} & \SI{144}{\milli\kelvin} \\
%JPA Physical Temperature & \SI{50}{\milli\kelvin} & \SI{25}{\milli\kelvin} \\
Total System Noise Temperature $T_\mathrm{sys}$\tnote{$\dagger$} & \SI{523}{\milli\kelvin} & \SI{341}{\milli\kelvin} \\
%Squeezing Speed-Up Factor & $1$ & $1.4$ \\
\textbf{Signal Combination (Sec.~\ref{sec:powercomb})} & & \\
Digital Phase Correction Error $\sigma_\phi$ & \SI{15}{\degree} & \SI{5}{\degree}  \\
Derived Power Combining Efficiency $\eta_c$ & \SI{93.7}{\percent} & \SI{99.3}{\percent}  \\
%\textbf{Instantaneous Data Taking} & & \\
%Time Spent Scanning & \SI{21}{\percent} & \SI{28}{\percent}  \\
\end{tabular}
\end{ruledtabular}
\begin{tablenotes}[flushleft]
\footnotesize
\item[$\dagger$] Frequency-dependent parameter. Its mean value across the experiment's bandwidth is quoted here.
\end{tablenotes}
\end{threeparttable}
\caption{\label{tab:parameters}%
Summary of critical design parameters for ADMX-EFR. For the frequency dependence of $Q_0$ and $C$, see Fig.~\ref{fig:cavity_performance_parameters}.
}
\end{table}
From these parameters the discovery potential of the overall design can be estimated by calculating the frequency-dependent instantaneous scan rate from Eqn.~\eqref{eq:instscanspeed}. The full frequency-dependent cavity and system noise parameters are taken into account.
The estimated scan rate for both \emph{threshold} and \emph{advanced} scenarios is shown in Fig.~\ref{fig:scanrate}.
The \emph{advanced} scan rate is about ten times higher than the \emph{threshold}.

Experiment downtime dictates the time available for scanning, affecting the overall scan rate. Accurately predicting downtime thus affects forecasts of discovery potential. To do so, the total experiment time is decomposed into a sum of contributions. Total experiment time $T_\mathrm{tot}$ can be expressed as

\begin{align}\label{eq:downtime}
    T_\mathrm{tot}=\tau_\mathrm{ops}+\tau_\mathrm{seek}+F_\mathrm{rescan}\left(\tau_\mathrm{scan}+T_\mathrm{scan}\right),
\end{align}

\noindent where $T_\mathrm{scan}$ is the time spent on digitization alone, and the downtime $\tau$ definitions and their calculated values in both the \emph{threshold} and \emph{advanced} scenarios are given in Tab.~\ref{tab:downtime}. 

\begin{table}[b]%The best place to locate the table environment is directly after its first reference in text
\begin{ruledtabular}
\begin{tabular}{lrr}
\textrm{Downtime Activity}&
\textrm{Threshold}&
\textrm{Advanced}\\
\colrule
\textbf{Operational} ($\tau_\mathrm{ops}$)& & \\
Initial Cooldown & 3~months & 1~month \\
Tuning Rod Exchange & 6~months & 3~months \\
\textbf{During Scanning} ($\tau_\mathrm{scan}$) &  &  \\
Cavity Tuning and Locking & 20~days & 20~days \\
Antenna Coupling Adjustment & 23~days & 12~days \\
JPA Rebiasing & 32~days & 16~days \\
\textbf{During Rescans} & & \\
Data Taking Multiplier ($F_\mathrm{rescan}$) & 1.1 & 1.1 \\
Rapid Tuning Rod Motion ($\tau_\mathrm{seek}$)& 166~days & 83~days \\
\textbf{Total Downtime} & 18.8~months & 8.6~months \\
\colrule 
\textbf{Overall Scan Time} ($F_\mathrm{rescan}T_\mathrm{scan})$ & 17.2~months & 27.4~months \\
\textbf{Total Time} ($T_\mathrm{tot})$ & 36~months & 36~months \\
\end{tabular}
\end{ruledtabular}
\caption{\label{tab:downtime}%
Summary of downtime activities during operation for ADMX-EFR. The rows, as indicated, correspond to terms in Eqn.~\eqref{eq:downtime}.
}
\end{table}

First, we discuss operational downtime, $\tau_\mathrm{ops}$, which encapsulates all time during which the experiment is not active. Based on the thermal performance discussed in Sec.~\ref{sec:magnet}, the tandem DRs can cool the experiment to base temperature in approximately one month. Without interruptions, this implies an additional three months is needed in the middle of the experiment to warm up, exchange the tuning rods and parapets to access higher frequencies, and cool down again, which we assume for the \emph{advanced} case. In the \emph{threshold} case, we account for the possibility of a critical failure during operations requiring another complete DR cycle, as well as an additional two months during the initial commissioning to account for setbacks before data taking.
%The \emph{objective} case assumes there are no unforeseen challenges, and is based solely on the fridge's turnaround time. The \emph{threshold} case allows for challenges requiring time for diagnostics to arise.

Second, we address the downtime incurred while the experiment is scanning across parameter space, $\tau_\mathrm{scan}$. This term is a sum of several contributions. 

The first of these is the time spent tuning the 18-cavity array by one frequency step and locking them to the target frequency. Because of the large number of cavities locking in parallel, we expect the locking time to be close to the worst-case time for any given step. Dedicated tests of a cryogenic \SI{4}{\giga\hertz} two-cavity system have found that the locking scheme takes at worst \SI{10}{\second}. Since there are 18 cavities in the array, we pessimistically assume that at least one cavity will require the worst-case locking time in the \emph{threshold} case, while the \emph{advanced} case allows for a factor of two improvement in the locking algorithm.
%{\color{blue} [TD: Will we discuss the locking scheme in more detail in the relevant section?]} 
The total time spent on locking is calculated assuming a frequency step of $\nu_0/10Q_L$. Since there are twice as many steps in the \emph{advanced} case, the total downtime remains the same.

The next source of downtime while scanning is adjusting the antenna depths for optimal coupling. During previous ADMX runs, this operation took \SI{1}{\second} per kHz of scanned parameter space. We take the historical precedent as the \emph{threshold} case. Heat load is typically the limiting factor on speed for cryogenic motors. The previous antenna adjustment system uses drive rods from hotter stages which introduce friction and passive heat loads. The \emph{advanced} case therefore increases the motor speed by a factor of two to account for the new piezoelectric antenna actuators which operate without those drive rods. This assumption is conservative, because dissipated heat as low as \SI{50}{\milli\joule} per degree has been recorded for rotational piezoelectric motors used for cryogenic actuation \cite{ADMX:2018ogs}.

The final source of downtime while scanning is JPA rebiasing. As the resonators tune to new frequencies, the narrow-band JPAs must be automatically rebiased to tune along with them. In ADMX's run 1D (ongoing at time of writing), the flux-pumped JPA is rebiased every \SI{20}{\kilo\hertz} on average, a procedure which takes \SI{75}{\second}. The total JPA rebias time in the \emph{threshold} case was calculated from the above numbers, scaling the rebias frequency step by a factor of three (to \SI{6.67}{\kilo\hertz}) to reflect the higher-frequency JPAs employed. Note that all 18 JPAs are biased simultaneously.
The JPA rebias procedure is a na\"ive grid search of pump tone parameter space which can be readily improved. An optimization algorithm requiring fewer samples, such as Nelder-Mead \cite{neldermead} or BOBYQA \cite{bobyqa}, would drastically reduce the time required to bias the JPAs. Making a modest assumption about these improvements, the JPA rebias downtime in the \emph{advanced} case is reduced by a factor of two.
%(accounting for the doubling of the number of JPA rebias operations required by the squeezing configuration).

Finally, we must account for the extra time taken to conclusively reject statistical outliers and RF interference. During these ``rescans,'' the resonant frequency is tuned back to candidate signals to acquire more data, observe enhancement on resonance and perform other checks. Assuming a Q-width region around each 3-$\sigma$ signal excess is rescanned, at most 7\% of the bandwidth would need to be scanned again due to statistics. We assume an additional 3\% of the bandwidth is occupied by transient RF interference. These assumptions are captured by $F_\mathrm{rescan}=1.1$, which multiplies all times associated with scanning (both digitization and downtime). We must also account for the non-trivial time required to move the tuning rods to the desired position, $\tau_\mathrm{seek}$. If we assume maximally inconvenient rescan targets, we must tune across the entire experiment bandwidth two additional times (a seek and a return). Note that this is separate from the cavity tuning and locking time during typical operation already included in $\tau_\mathrm{scan}$. 
Previous ADMX operations recorded a maximum tuning rate during rescans of 0.4~MHz/h near 0.8~GHz, corresponding to a rate of 1~MHz/h at 2~GHz, assuming tuning rate is relative to frequency.
The scaled tuning rate is used for the \emph{threshold} case in this estimate. 
That tuning rate is doubled in the \emph{advanced} case for the same reason given in the above discussion regarding the antenna adjustment motors.
%The \emph{objective} estimate calculates the tuning rate by assuming the rotary piezoelectric motor speed is limited by fridge cooling power, using known heat dissipated per degree and number of degrees each of the 18 motors must sweep. {\color{blue} TD: Cite Christian sidecar 2019 paper with the rotary piezo heating number?} 

%{ \color{blue} add plots showing the scan speed efficiency of current runs? }
%Assuming similar performance of ADMX-EFR compared to previous ADMX runs, we conclude that the actual scan speed is $20\,\%$ (\emph{Threshold}) or $30\,\%$ (\emph{Objective}) of the instantaneous scan speed.

With a total experiment run time of three years accounting for all downtime discussed above, DFSZ axion-photon coupling sensitivity is achieved in the \textit{threshold} case. With \textit{advanced} assumptions, sensitivity reaches 2.2 times DFSZ in that time frame, or reaches DFSZ sensitivity to dark matter density as low as \SI{0.1}{\giga\electronvolt\per\centi\meter\cubed}.

\begin{figure*}
    \centering
    \includegraphics[width=\linewidth]{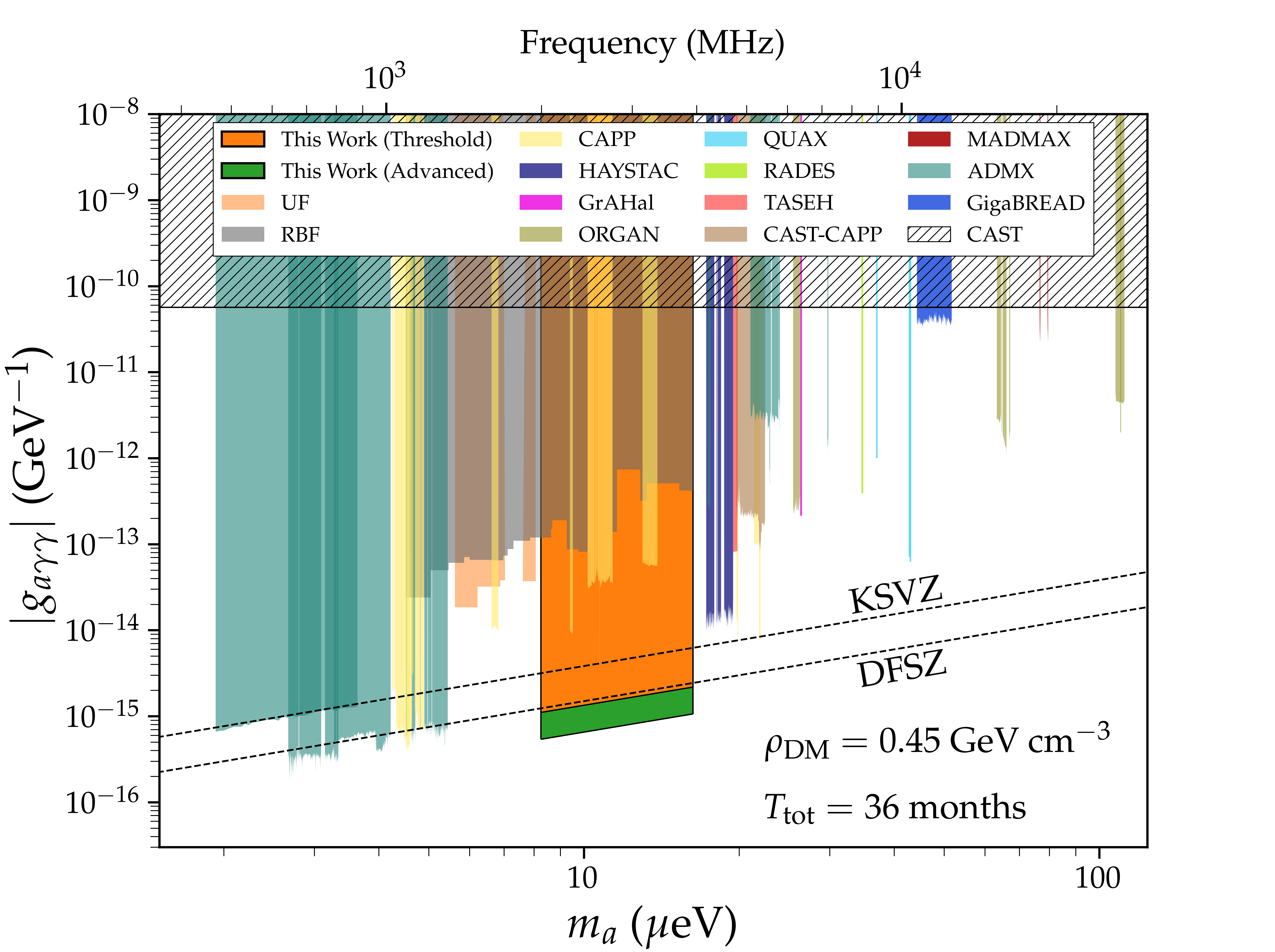}
    \caption{\label{fig:sensitivity} Projected sensitivity in $|g_{a\gamma\gamma}|$ as function of axion mass for ADMX-EFR over a three-year period given both \emph{threshold} (orange) and \emph{advanced} (green) design parameters (see Tab.~\ref{tab:parameters}). Note that there are no gaps in sensitivity due to mode crossings because the array features cavities of three different lengths (see Sec.~\ref{sec:cavity} for details). KSVZ and DFSZ coupling strengths are shown as dashed lines. The dark matter density is assumed to be \SI{0.45}{\giga\electronvolt\per\centi\meter\cubed}. 
    Limits set by ADMX are shown in teal \cite{PhysRevLett.104.041301,ADMX:2018gho,ADMX:2019uok,ADMX:2021nhd,bartram2024axiondarkmatterexperiment,ADMX:2021mio,ADMX:2018ogs,Carosi_2025}. The other limits shown are those set by University of Florida (UF)~\cite{uf_limit_1990}, Rochester-Brookhaven-Florida (RBF)~\cite{Wuensch:1989sa}, Center for Axion and Precision Physics (CAPP)~\cite{Lee:2020cfj,Jeong:2020cwz,CAPP:2020utb,Lee:2022mnc,Kim:2022hmg,Yi:2022fmn,Yang:2023yry,Kim:2023vpo,CAPP_prx}, Haloscope At Yale Sensitive To Axion Cold dark matter (HAYSTAC)~\cite{HAYSTAC:2018rwy,Backes:2021,HAYSTAC:2023cam}, Grenoble Axion Haloscope project (GrAHal)~\cite{Grenet:2021vbb}, Oscillating Resonant Group AxioN experiment (ORGAN)~\cite{McAllister:2017lkb,Quiskamp:2022pks,Quiskamp:2023ehr}, MAgnetized Disc and Mirror Axion eXperiment (MADMAX)~\cite{c749-419q}, QUaerere AXions experiment (QUAX)~\cite{Alesini:2019ajt,Alesini:2020vny,Alesini:2022lnp,QUAX:2023gop,QUAX:2024fut}, Relic Axion Dark Matter Exploratory Setup (RADES)~\cite{CAST:2020rlf}, Taiwan Axion Search Experiment with Haloscope (TASEH)~\cite{TASEH:2022vvu}, Gigahertz Broadband Reflector Experiment for Axion Detection (GigaBREAD)~\cite{GigaBREAD:2025lzq}, CAST-CAPP~\cite{Adair:2022rtw}, and CERN Axion Solar Telescope (CAST)~\cite{cast2024}. Figure produced with plotting code based on \cite{AxionLimits}.}
\end{figure*}

\section{\label{sec:conclusion}Conclusion}
% Stefan
%In this paper we have introduced the design of ADMX-EFR, an array of 18 cavities in a \SI{9.4}{\tesla} magnetic field to cover the frequency range from $2$ to \SI{4}{\giga\hertz}. We showed that based on ... can be acheived.

\noindent
In this paper we have introduced the design and detailed sensitivity estimate for ADMX-EFR, a baseline concept for a resonator array haloscope using existing and near-term technologies that can probe the axion-photon coupling to the DFSZ limit from 2 to \SI{4}{\giga\hertz} (axion masses 8.27 to~\SI{16.54}{\micro\electronvolt}). It uses 18 independently tuned cavities each with their own full RF receiver chains, including parametric amplifiers. For a full overview of ADMX-EFR's design parameters, see Tab.~\ref{tab:parameters}. 

The sensitivity of ADMX-EFR over three years of operation has been calculated. This calculation involves detailed simulations of the cavity and RF receiver chain to extract the haloscope's instantaneous scan rate, as well as a detailed analysis of the time required to perform each task ancillary to digitizing (see Tab.~\ref{tab:downtime}). We project that ADMX-EFR would reach DFSZ sensitivity across its entire tuning range in the \emph{threshold} case, which assumes performance extrapolated from demonstrated components. In the \emph{advanced} case, where technologies are developed further for use in ADMX-EFR, the experiment could probe couplings over two times below the DFSZ benchmark, or, conversely has DFSZ sensitivity assuming local $\rho_\mathrm{DM}$ as low as \SI{0.1}{\giga\electronvolt\per\centi\metre^3}. ADMX-EFR's projected scan rate is greater than previous ADMX runs in the University of Washington facility primarily because of the larger volume and higher-field magnet, which compensate for the elevated quantum noise inherent to higher frequencies. The speed-up is also facilitated by the development of high-gain and near-quantum-limited JPAs at the relevant frequencies. At the same time, the need for digital phase correction and cavity locking to gain the full benefit of a resonator array (rather than the same number of independent resonators) dramatically increases operational complexity. Noise calibration and simultaneous operation of 18 JPAs, as well as reliable adjustment of 36 independent actuators for resonator tuning and antenna coupling, is challenging to accomplish in practice. Further, the parapet design allowing for cold electronics configuration changes without a full warm-up, critical to maintaining low operational downtime for the system, has not been demonstrated.

Recall that these measures offset serious disadvantages of the resonator array haloscope: the independent RF chains enable digital phase correction and thus allow $\mathrm{ SNR}\propto N$  (see Sec.~\ref{sec:powercomb}).

At the time of writing, there is no large-scale experiment planned to probe the axion above \SI{2}{\giga\hertz}. This study explores in detail one potential avenue for DFSZ axion discovery in this region of parameter space. In particular, it illustrates the value of a large-bore, high-field magnet for large-scale axion experiments. ADMX-EFR cavities are approximately \SI{1}{\metre} long, while the magnet measures \SI{3}{\metre} along its axis. This allows other haloscope designs to be tested in the remaining volume, demonstrating new technology while achieving competitive axion sensitivity.

This study estimates the sensitivity of a cavity array haloscope using demonstrated or near-term technologies. The performance predicted here can be used as a benchmark to compare other emerging high-frequency haloscope designs. 
%The authors also hope that the rigorous framework presented here may be replicated to accurately assess the performance of proposed axion dark matter search experiments. \td{Too aggrandizing?}

\section*{Acknowledgments}
\noindent
This work was supported by the U.S. Department of Energy through Grants No. DE-SC0009800, No. DESC0009723, No. DE-SC0010296, No. DE-SC0010280, No. DE-SC0011665, No. DEFG02-97ER41029, No. DEFG02-96ER40956, No. DEAC52-07NA27344, No. DEC03-76SF00098, No. DE-SC-0022148, and No. DE-SC0017987. Fermilab is a U.S. Department of Energy, Office of Science, HEP User Facility. Fermilab is managed by Fermi Research Alliance, LLC (FRA), acting under Contract No. DE-AC02-07CH11359. PNNL is a multi-program national laboratory operated for the U.S. DOE by Battelle Memorial Institute under Contract No. DE-AC05-76RL01830. Additional support was provided by the Heising-Simons Foundation and by the Lawrence Livermore National Laboratory and Pacific Northwest National Laboratory LDRD offices. C. Boutan is supported by DOE Office of Science, High Energy Physics, Early Career Award (No. FWP 77794 at PNNL). C. Bartram and A. Yi are supported by the Department of Energy, Laboratory Directed Research and Development program at SLAC National Accelerator Laboratory, under contract DE-AC02-76SF00515 and as part of the Panofsky Fellowship awarded to C. Bartram. J. C. acknowledges support from the U.S. Department of Energy, Office of Science, National Quantum Information Science Research Centers. E. J. D., C. M., and M. G. P. acknowledge support from the UK Science and Technology Facilities Council (STFC) under Grants No. ST/T006811/1 and No. ST/X005879/1. M. G., E. H., A. Q., M. E. T., and B. T. M.’s participation is funded by the ARC Centre of Excellence for Engineered Quantum Systems and No. CE170100009, Dark Matter Particle Physics, No. CE200100008. LLNL Release No. LLNL-JRNL-2026833-DRAFT. LANL Release No. LA-UR-XX-XXXX

\bibliography{refs}% Produces the bibliography via BibTeX.

\end{document}